# Evidence for an external origin of heavy elements in hot DA white dwarfs

M.A. Barstow¹, J.K. Barstow¹,², S.L. Casewell¹, J.B. Holberg³ and I. Hubeny⁴

*¹ Department of Physics and Astronomy, University of Leicester, University Road, Leicester LE1 7RH, UK*
*² Department of Astrophysics, University of Oxford, Denys Wilkinson Building, Keble Road, Oxford, OX1 3RH*
*³ Lunar and Planetary Laboratory, Sonett Space Science Building, University of Arizona, Tucson, AZ 85721, USA*
*⁴ Steward Observatory, The University of Arizona, 933 N. Cherry Ave, Tucson, AZ 85721, USA*

**ABSTRACT**

We present a series of systematic abundance measurements for 89 hydrogen atmosphere (DA-type) white dwarfs with temperatures spanning 16000-77000K drawn from the *FUSE* spectral archive. This is the largest study to date of white dwarfs where radiative forces are significant, exceeding our earlier work, based mainly on *IUE* and *HST* data, by a factor three. Using heavy element blanketed non-LTE stellar atmosphere calculations, we have addressed the heavy element abundance patterns making completely objective measurements of abundance values and their error ranges using a $\chi^2$ fitting technique. We are able to establish the broad range of abundances seen in a given temperature range and establish the incidence of stars which appear, in the optical, to be atmospherically devoid of any material other than H. We compare the observed abundances to predictions of radiative levitation calculations, revealing little agreement. We propose that the supply of heavy elements is accreted from external sources rather than being intrinsic to the star. These elements are then retained in the white dwarf atmospheres by radiative levitation, a model that can explain both the diversity of measured abundances for stars of similar temperature and gravity, including cases with apparently pure H envelopes, and the presence of photospheric metals at temperatures where radiative levitation is no longer effective.

**Keywords:** Stars: white dwarfs -- abundances -- ultraviolet:stars.

## 1 INTRODUCTION

White dwarfs are the end points of the evolution of all stars with masses less than approximately 8M$_\odot$. During their lives, the parent star may have been a component of a binary and/or hosted a planetary system. There are likely to have also been encounters with interstellar clouds, affecting the local environment. Furthermore, there is now strong evidence that planetary debris disks, from which atmosphere polluting material is being accreted, surround many white dwarfs. Hence, there is the exciting prospect that the study of the composition of white dwarf atmospheres can provide information on the bulk composition of extra-solar planets. Such studies have concentrated on the relatively simple situation of white dwarfs towards the cool end of the temperature range, below about 20,000K, where metals should sink out of their atmospheres within a few days, demonstrating that there must be a continuous supply of material. Above this temperature, the picture is more complex, as radiation pressure can act against the downward diffusion due to gravity. Therefore, observed metal abundances are more difficult to interpret. How much arises from radiative levitation of material already contained within the stellar envelope and what proportion, if any, is attributable to an external source?

  Measurements of the physical parameters of white dwarf stars are crucial to understanding their evolution and their relationship with progenitor systems, particularly through their evolution on the giant branch. Of particular importance are accurate measurements of photospheric composition and structure, which can tell us how the He and heavy element content of white dwarfs may change as the stars cool. The existence of two distinct groups of hot white dwarfs, having either hydrogen-rich or helium-rich atmospheres is understood to arise from the number of times the progenitor star ascends the red giant branch and the amount of H (and He) lost through the successive phases of mass-loss. Consequently, it seems clear that each group descends from their respective proposed progenitors, the H-rich and He-rich central stars of planetary nebulae (CSPN). Nevertheless, the large preponderance of hot ($T_{eff}$ > 20000K) H-rich DA white dwarfs over the He-rich DOs by factors of 6 to 21 (depending on the temperature) observed in the Sloan Digital Sky Survey (SDSS, Krzesinski et al. 2009), compared to the relative number (4:1) of H- and He-rich Central Stars of Planetary Nebulae (Napiwotzki 1999), is hard to explain. On the other hand, the improved white dwarf population statistics produced by the SDSS (Eisenstein et al. 2006; Kleinman et al. 2013) have demonstrated that an apparent gap in the He-rich track between ≈45000K and 30000K between the hot DO and cooler DB white dwarfs, identified by Wesemael et al. (1985) and Liebert et al. (1986), is not absolute, but rather contains relatively few stars.

In understanding white dwarf evolution, we strive to accurately measure a number of physical parameters for each star. The effective temperature ($T_{eff}$) establishes the star's position on its evolutionary path and is conventionally obtained by comparing the Balmer line profiles in optical spectra with synthetic spectra from stellar model atmosphere calculations (e.g. Bergeron, Saffer & Liebert 1992). In some cases, the Balmer lines may not be accessible, for example if the white dwarf has an unresolved, much brighter, main sequence companion, i.e. 'Sirius-Like Systems' (see Holberg et al. 2013). Then, a similar technique can be applied to the Lyman series lines (e.g. Barstow et al. 2003a). The reliability of the work depends on the assumption that the line profile techniques are unbiased estimators of $T_{eff}$ in all cases. However, several authors have demonstrated that treatment of details in the stellar atmosphere calculations can affect the outcome in a number of ways. For example, Barstow, Hubeny & Holberg (1998) showed that the presence of substantial blanketing from photospheric heavy elements significantly alters the Balmer line profiles at a given effective temperature. Hence, the temperature scale of the hottest, most metal-rich DA white dwarfs, must at least take into account the photospheric composition of each star. In addition, there remains an unsolved discrepancy between the inferred temperatures from Balmer and Lyman line work in the hottest DA white dwarfs, where Lyman values for the hottest stars are systematically higher than the Balmer results (Barstow et al. 2003a). This problem is even more extreme in the DAO white dwarfs (H dominated atmospheres with detectable He) where the differences in the temperature values approach 50 per cent compared to 10-15 per cent in the DAs, in some cases (Good et al. 2004). Thus, it is apparent that there are as-yet undetermined additional effects that need to be taken into account in the stellar atmosphere models.

Our understanding of the composition of white dwarfs has evolved radically during the past decades. Schatzman (1958) pointed out that the strong gravitational field (log $g$ ~ 8 at the surface) of a white dwarf would cause rapid downward diffusion of elements heavier than the principal H or He components, yielding very pure atmospheres devoid of any other material. Hence, the detections of heavier elements in *IUE* echelle spectra (e.g. Dupree and Raymond 1982) were initially interpreted as arising from circumstellar material. Eventually, determination of the photospheric radial velocities of several stars, coupled with an *EXOSAT* EUV transmission-grating spectrum of Feige 24 (Vennes et al. 1988), established that the observed features were photospheric in origin. The physical explanation was that the heavy elements were prevented from sinking out of the atmosphere by radiation pressure (e.g. Vauclair, Vauclair & Greenstein 1978; Chayer et al. 1995), a mechanism termed "radiative levitation". However, with few UV spectra available, it was far from clear whether stars like Feige 24 were typical of the DA population or if stars with pure H atmospheres were more representative.

Ultimately, the larger statistical sample (~100 stars) of the *ROSAT* Wide Field Camera EUV sky survey revealed that the EUV and soft X-ray luminosities of white dwarfs hotter than about 50000 K are much lower than expected, while cooler stars have luminosities consistent with pure H envelopes (Barstow et al. 1993; Marsh et al. 1997). These results were consistent with the idea that radiative levitation determines the photospheric composition of the hottest white dwarfs, injecting heavy elements that suppress the EUV/soft X-ray flux, but the broadband *ROSAT* photometric data were not able to determine what elements were contributing to the opacity. However, prior to the *ROSAT* survey a number of far-UV spectra had been obtained by *IUE*, from which contributing elements and their abundances could be determined. Subsequently, further spectra were obtained by both *IUE* and with the *HST* GHRS and STIS instruments in follow-up studies. The resulting sample of 25 hot white dwarfs represented the largest study of hot DA white dwarf compositions to date (Barstow et al. 2003b).

These far UV spectra provide detections of C, N, O, Si, Fe and Ni in the hottest white dwarfs, qualitatively consistent with detections expected from radiative levitation calculations (e.g. Chayer, Fontaine & Wesemael 1995; Chayer et al. 1994, 1995). However, the measured abundances do not match the predicted values at all well. A limitation of the Barstow et al. (2003b) sample is its strong bias to the higher temperatures: those stars expected to have photospheric heavy elements and, therefore, those most "interesting" for spectroscopic observations. At temperatures below 50000 K, most stars exhibit the presence of photospheric heavy elements but a few do not (7 detections, 2 non-detections). It is not possible to draw any conclusions about the general abundance pattern at these temperatures from the relatively small number of these objects. Hence, even the qualitative agreement with the prediction, by the radiative levitation calculations, that particular elements should be present at certain temperatures may well be just a selection bias.

None of the detailed theoretical studies of radiative levitation in white dwarf stars (see Chayer, Fontaine & Wesemal 1995; Chayer et al. 1994, 1995; Schuh, Dreizler & Wolff 2002) consider the possible source or the size of the reservoir of heavy elements. The predicted element abundances are based on an assumption that equilibrium conditions occur with radiative forces being balanced by gravity and that as much material as is necessary to achieve this is available. Furthermore, in general, no other mechanisms, such as accretion or the possibility of mass-loss are taken into account.

Accretion from the interstellar medium (ISM) has been invoked as a mechanism to explain the presence of heavy elements in some white dwarfs, particularly where radiative effects are expected to be negligible, in the cooler DA stars, or observed abundances are anomalously high. For example, Dupuis, Fontaine & Wesemael (1993) show good agreement between a two-phase accretion model and the observed abundance measurements available at the time. However, several factors challenge the ISM pollution scenario. The He dominated DZ stars exhibit a dearth of hydrogen relative to Ca, suggesting any accretion must be from volatile-depleted material (e.g. Dufour et al. 2007). In addition, it has been established that diffusion timescales for heavy elements are as short as a few days in the H-rich DAZ white dwarfs. Thus, with galactic locations far from known interstellar clouds and material accreted from the ISM will have long-since settled out of the atmosphere and another source of material is required (Koester et al. 2005; Zuckerman et al. 2003).

In the last few years an increasing number of heavy element-contaminated white dwarfs have been discovered, associated with infrared excesses and closely orbiting rings of dusty and gaseous debris (Farihi, Jura & Zuckerman 2009; Jura, Farihi & Zuckerman 2009; Gäensicke et al. 2008). Such debris offers a ready source of material for accretion into the atmospheres of the host stars. A statistical analysis of the galactic position and kinematics of DZ and DAZ stars from the SDSS reported no correlation between accreted Ca abundances and spatial–kinematical distributions relative to interstellar material (Farihi et al. 2010a). Furthermore, they find that the DZ and DC stars belong to the same population, with similar basic atmospheric compositions, effective temperatures, spatial distributions and Galactic space velocities. Accretion from the ISM cannot simultaneously account for the polluted and non-polluted sub-populations. Therefore, it is probable that circumstellar matter, the rocky remnants of terrestrial planetary systems, contaminates these white dwarfs.

If, as seems likely, white dwarfs are accreting material from extra-solar planetary debris, there is the possibility of determining the composition of these bodies from the compositions of the stellar photospheres (see e.g. Jura et al. 2009, 2012; Gäensicke et al. 2012). It is clear from both earlier and more recent studies that the photospheric composition of white dwarfs arises from a complex interplay of physical effects. At the extremes of the white dwarf temperature range one mechanism may dominate: for example, radiative levitation in hot stars and accretion in the coolest. However, in most objects we will need to understand the relative contributions from several effects. Debris disks, as a source of accreted material, are readily detectable, if present, at the cool end of the cooling sequence. These disks are believed to have formed from tidal disruption of their parent bodies and consist of small solids. They are entirely within the Roche limit of the star (approximately 1 $R_\odot$, see e.g. Jura et al. 2007 and Farihi et al. 2009). These solids are brought inward by Poynting-Robertson drag (Rafikov 2011) until they sublimate. The metals will then accrete via the magneto-rotational instability.

The correspondence between metals and debris disks is far from one-to-one. For example Zuckerman et al. (2003) find approximately 25 per cent of cool of DA stars can be classified DAZ in the optical but only about 3 per cent of DAZ stars show the unambiguous presence of debris disks in the IR. For hot white dwarfs, where metals seen in the UV may be supported by radiative equilibrium, there have been no detections of similar debris disks in the IR. This may be simply because, at higher temperatures, the planetary debris is sublimated into gaseous disks. von Hippel et al. (2007) calculate that no solids would survive within the Roche limit of a white dwarf for a stellar temperature above 24000K or 32000K, depending on assumptions about disk geometry and sublimation temperature of dust grains.

The *FUSE* archive contains a large sample of hot DA white dwarf observations numbering more than 100 stars and spanning a range of effective temperatures similar to the *HST/IUE* sample of Barstow et al. (2003b). However, as the data were obtained for a variety of purposes, including study of the interstellar medium as well as white dwarf atmospheres, the full temperature range is sampled more uniformly by the *FUSE* data. The different wavelength range (912-1180 Å) covered by *FUSE* reveals a slightly different group of ionic species than either *HST* or *IUE*. Nevertheless, this white dwarf sample presents an opportunity to survey the composition of hot DA white dwarfs with a greater level of completeness and much better detail than hitherto.

We present the results of a detailed study of the composition of 89 white dwarfs where the spectra are of high enough signal-to-noise to allow abundance measurements with sensible uncertainties to be made, a sample more than a factor of 3 larger than that of Barstow et al. (2003b). We determine the range of abundances seen in a given temperature range and establish the incidence of stars which appear to be completely devoid of any material other than H in their atmospheres. We then compare the observed abundances to predictions of radiative levitation calculations.

## 2 OBSERVATIONS AND DATA REDUCTION

### 2.1 *FUSE* spectra

The *FUSE* mission was launched in 1999 and was successfully operated for 8 years, until 2007. Overviews of the *FUSE* mission, instrument design and in-orbit performance have been described in several papers, including Moos et al. (2000) and Sahnow et al. (2000). Briefly, the spectrographs are based on a Rowland circle design, comprising four separate optical paths (or channels). The channels must be co-aligned so that light from a single target properly illuminates all four channels, to maximize the throughput of the instrument. Spectra from the four channels are recorded on two microchannel plate detectors, with a SiC and LiF spectrum on each. Each detector is divided into two functionally independent segments (A and B), separated by a small gap. Consequently, there are eight detector segment/spectrometer channel combinations to be dealt with in reducing the data. Maintaining the co-alignment of individual channels was difficult in-orbit, mainly due to thermal effects. A target may completely miss an aperture for the whole or part of an observation, while being well centered in the others. Hence, in any given observation, not all of the channels may be available in the data. To minimize this problem, most observations have been conducted using the largest aperture available (LWRS, 30 x 30 arcsec). This limits the spectral resolution to between 15000 and 20000 for early observations or 23000 later in the programme when the mirror focusing had been adjusted (Sahnow et al. 2000), compared to the 24000-30000 expected for the 1.25 x 20 arcsec HIRS aperture.

The majority of the spectra utilized for the current study have been included in an earlier paper concerning O VI in the interstellar medium (Barstow et al. 2010). The initial sample of 95 objects was reduced to 80, after eliminating He-rich objects, stars contaminated with interstellar $H_2$ and cooler DA stars, where the spectral region containing the O VI features was too low signal-to-noise. In this work, the most important features are at longer wavelengths than the O VI lines, where the continuum flux in the cooler objects is not as severely attenuated by the broad Lyman lines. Therefore, we can include a few more of the cooler DAs, yielding a sample of 89 stars. All the white dwarfs included in our analysis are listed in Table 1, along with their physical parameters ($T_{eff}$, log $g$ from Barstow et al. 2010 together with the cooling age calculated from these). The stars are ordered by white dwarf RA/Dec number, where one has been assigned in the McCook & Sion (1999) catalogue, and any principal alternative name in general use is also given. Where there is no such name, the designation of the primary discovery survey is noted instead. Table 2 lists the datasets used for the white dwarf spectra studied in this paper.

**Table 1.** Summary of white dwarfs targets and their physical parameters.

| WD No. | Alt. Name/Cat* | $T_{eff}$ | log $g$ | Cooling Age Myr | WD No. | Alt. Name/Cat* | $T_{eff}$ | log $g$ | Cooling Age Myr |
|---|---|---|---|---|---|---|---|---|---|
| 0001+433 | REJ, EUVEJ | 46205 | 8.85 | 30.00 | 1337+701 | EG102 | 20435 | 7.87 | 56.00 |
| 0004+330 | GD2 | 47936 | 7.77 | 2.10 | 1342+442 | PG | 66750 | 7.93 | 0.95 |
| 0027-636 | REJ, EUVEJ | 60595 | 7.97 | 1.20 | 1440+753 | REJ, EUVEJ | 42400 | 8.54 | 11.00 |
| 0041+092 | BD+08 102 | 22113 | 7.71 | 35.00 | 1528+487 | REJ, EUVEJ | 46230 | 7.70 | 2.30 |
| 0050-332 | GD659 | 34684 | 7.89 | 6.10 | 1550+130 | NN Ser | 39910 | 6.82 | 0.69 |
| 0106-358 | GD683 | 28580 | 7.90 | 12.00 | 1603+432 | PG | 36257 | 7.85 | 5.30 |
| 0131-164 | GD984 | 44850 | 7.96 | 2.80 | 1611-084 | REJ, EUVEJ | 38500 | 7.85 | 4.30 |
| 0147+674 | REJ, EUVEJ | 30120 | 7.70 | 9.40 | 1615-154 | G153-41 | 38205 | 7.90 | 4.60 |
| 0226-615 | HD15638 | 52301 | 7.76 | 1.90 | 1620-391 | CD-38 10980 | 24760 | 7.92 | 23.10 |
| 0229-481 | REJ, EUVEJ | 63400 | 7.43 | 0.70 | 1620+647 | PG | 30184 | 7.72 | 9.30 |
| 0232+035 | Feige 24 | 62947 | 7.53 | 0.85 | 1631+781 | REJ, EUVEJ | 44559 | 7.79 | 2.70 |
| 0235-125 | PHL1400 | 32306 | 8.44 | 33.00 | 1635+529 | HD150100 | 20027 | 8.14 | 100.00 |
| 0236+498 | REJ, EUVEJ | 33822 | 8.47 | 28.00 | 1636+351 | 1638+349 | 36056 | 7.71 | 4.90 |
| 0252-055 | HD18131 | 30355 | 7.26 | 11.00 | 1648+407 | REJ, EUVEJ | 37850 | 7.95 | 4.80 |
| 0310-688 | LB3303 | 16181 | 8.06 | 180.00 | 1711+668 | REJ, EUVEJ | 60900 | 8.39 | 0.95 |
| 0320-539 | LB1663 | 32860 | 7.66 | 6.90 | 1725+586 | LB335 | 54550 | 8.49 | 1.30 |
| 0346-011 | GD50 | 42373 | 9.00 | 59.00 | 1734+742 | 29 Dra | 28795 | 8.00 | 12.00 |
| 0353+284 | REJ, EUVEJ | 32984 | 7.87 | 7.20 | 1800+685 | KUV | 43701 | 7.80 | 2.80 |
| 0354-368 | | 53000 | 8.00 | 1.60 | 1819+580 | REJ, EUVEJ | 45330 | 7.73 | 2.50 |
| 0416+402 | KPD | 35227 | 7.75 | 5.40 | 1844-223 | REJ, EUVEJ | 31470 | 8.17 | 19.00 |
| 0455-282 | REJ, EUVEJ | 58080 | 7.90 | 1.30 | 1845+683 | KUV | 36888 | 8.12 | 8.60 |
| 0457-103 | 63 Eri | 22261 | 7.30 | 17.00 | 1917+509 | REJ, EUVEJ | 33000 | 7.90 | 7.30 |
| 0501+524 | G191-B2B | 57340 | 7.48 | 1.10 | 1921-566 | REJ, EUVEJ | 52946 | 8.16 | 1.60 |
| 0512+326 | HD33959C | 42849 | 8.06 | 2.60 | 1942+499 | REJ, EUVEJ | 33500 | 7.86 | 6.80 |
| 0549+158 | GD71 | 32780 | 7.83 | 7.30 | 1950-432 | HS | 41339 | 7.85 | 3.40 |
| 0603-483 | REJ, EUVEJ | 33040 | 7.80 | 7.00 | 2000-561 | REJ, EUVEJ | 44456 | 7.54 | 2.40 |
| 0621-376 | REJ, EUVEJ | 62280 | 7.22 | 0.41 | 2004-605 | REJ, EUVEJ | 44200 | 8.14 | 3.50 |
| 0659+130 | REJ0702+129 | 39960 | 8.31 | 7.70 | 2011+398 | REJ, EUVEJ | 47057 | 7.74 | 2.20 |
| 0715-704 | REJ, EUVEJ | 44300 | 7.69 | 2.60 | 2014-575 | LS210-114 | 26579 | 7.78 | 16.00 |
| 0802+413 | KUV | 45394 | 7.39 | 1.80 | 2020-425 | REJ, EUVEJ | 28597 | 8.54 | 69.00 |
| 0809-728 | REJ, EUVEJ | 30585 | 7.90 | 8.90 | 2032+248 | Wolf 1346 | 19150 | 7.91 | 77.20 |
| 0830-535 | REJ, EUVEJ | 29330 | 7.79 | 11.00 | 2043-635 | BPM13537 | 25971 | 8.36 | 68.00 |
| 0905-724 | HR3643 | 21551 | 7.47 | 25.00 | 2111+498 | GD394 | 38866 | 7.84 | 4.20 |
| 0937+505 | PG | 35552 | 7.76 | 5.30 | 2116+736 | KUV | 54486 | 7.76 | 1.40 |
| 1019-141 | REJ, EUVEJ | 31340 | 7.79 | 11.00 | 2124-224 | REJ, EUVEJ | 48297 | 7.69 | 2.00 |
| 1021+266 | HD90052 | 35432 | 7.48 | 4.30 | 2124+191 | HD204188 | 33290 | 8.90 | 110.00 |
| 1024+326 | REJ, EUVEJ | 41354 | 7.59 | 3.00 | 2146-433 | REJ, EUVEJ | 67912 | 7.58 | 0.69 |
| 1029+537 | REJ, EUVEJ | 44980 | 7.68 | 2.50 | 2152-548 | REJ, EUVEJ | 45800 | 7.78 | 2.40 |
| 1040+492 | REJ, EUVEJ | 47560 | 7.62 | 2.00 | 2211-495 | REJ, EUVEJ | 65600 | 7.42 | 0.61 |
| 1041+580 | REJ, EUVEJ | 29016 | 7.79 | 11.00 | 2257-073 | HD217411 | 38010 | 7.84 | 4.50 |
| 1057+719 | PG | 39555 | 7.66 | 3.50 | 2309+105 | GD246 | 51300 | 7.91 | 1.70 |
| 1109-225 | HD97277 | 36885 | 7.40 | 4.00 | 2321-549 | REJ, EUVEJ | 45860 | 7.73 | 2.40 |
| 1234+481 | PG | 55570 | 7.57 | 1.30 | 2331-475 | REJ, EUVEJ | 56682 | 7.64 | 1.20 |
| 1254+223 | GD153 | 39390 | 7.77 | 4.60 | 2350-706 | HD223816 | 76690 | 7.83 | 0.92 |
| 1314+293 | HZ43 | 49435 | 7.95 | 1.90 | | | | | |

* PG = Palomar Green, HS = Hamburg Schmidt, REJ = ROSAT EUV, EUVEJ = EUVE, KUV = Kiso UV, KPD = Kitt Peak Downes

## 2.2 Data reduction

All the data used in this work (see Table 2) were obtained from the Mikulski Archive for Space Telescopes (MAST) and were subject to reprocessing using version 3.2 of the CALFUSE pipeline. The data supplied by the archive consist of a set of separate exposures for each of the 8 channel/detector segment combinations. The wavelength binning (0.006 Å) oversamples the true spectral resolution by a factor 10. Therefore, we have established a reliable procedure to combine and resample all the spectra to optimize the bin size, with respect to the spectral resolution, and the signal-to-noise (see Barstow et al. 2003b; Barstow et al. 2010), resulting in a single composite spectrum spanning the full *FUSE* wavelength range for each star. An example of the resulting spectra, taking one of the higher signal-to-noise data sets is shown in Fig. 1. The region of poor signal-to-noise seen in the 1080 -1087 Å region is due to the absence of data from the higher effective area LiF grating at those wavelengths.

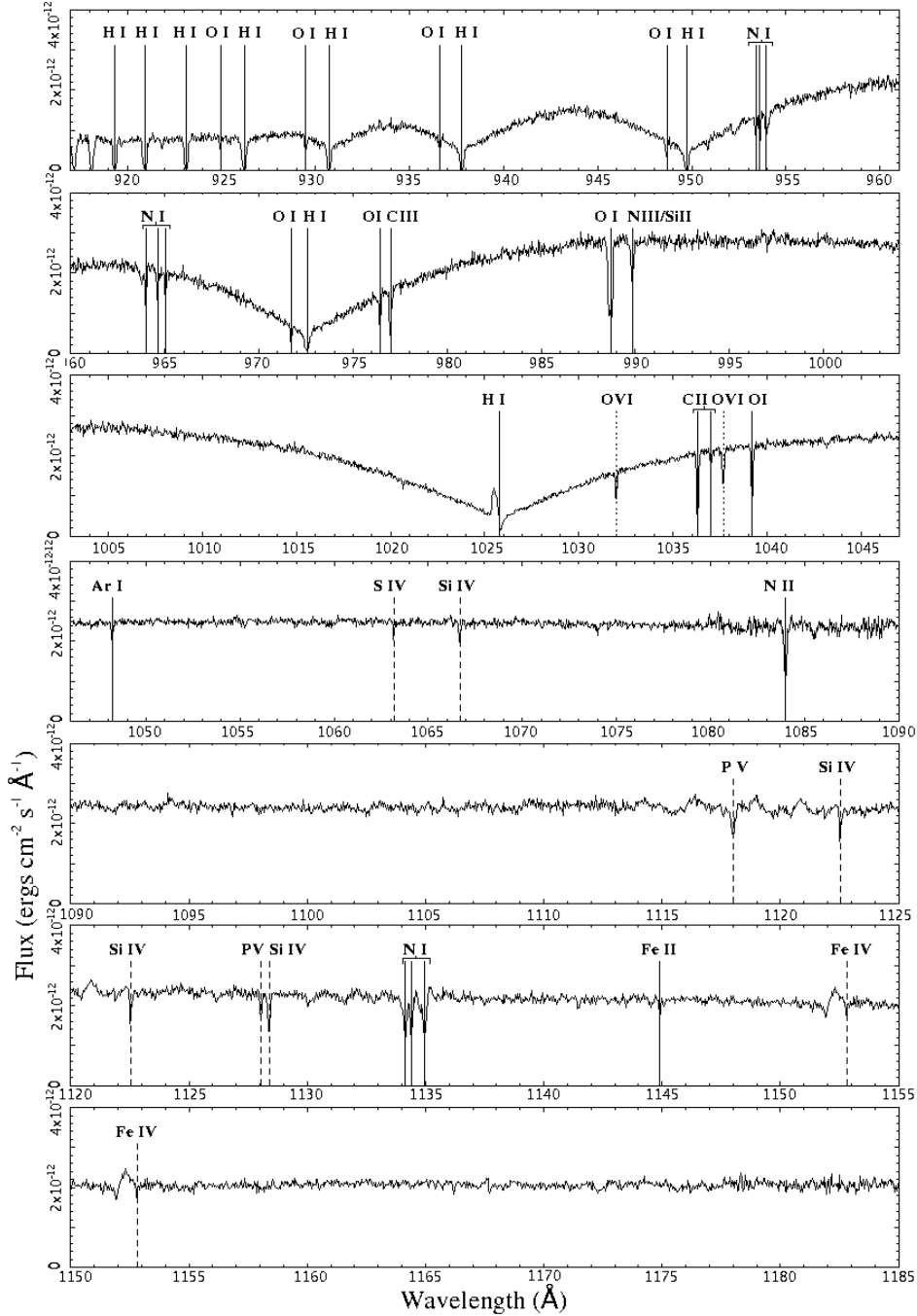

**Figure 1.** *FUSE* spectrum of the hot DA white dwarf WD 0131-164 (GD 984), with the main photospheric and interstellar features labelled. The wavelengths of all these absorption lines can be obtained from Table 3. We note that the line identified as photospheric Si IV 1066.62Å in this plot may often be contaminated by interstellar Ar I, which occurs at the same wavelength.

**Table 2.** Summary of *FUSE* observations. The last columns (*) list the G191-B2B observations included in the co-added spectrum

| WD No. | Observation | Start Time | WD No. | Observation | Start Time | Observation* | Start Time |
|---|---|---|---|---|---|---|---|
| 0001+433 | E90305010 | 24/11/2004 19:34 | 1254+223 | M10104020 | 29/04/2000 21:32 | M10102010 | 13/10/1999 01:25 |
| 0004+330 | P20411020-40 | 14/09/2002 10:07 | 1314+293 | P10423010 | 22/04/2000 21:19 | M10305010 | 12/11/1999 07:35 |
| 0027-636 | Z90302010 | 30/09/2002 15:44 | 1337+701 | B11903010 | 05/05/2001 06:43 | M10305020 | 20/11/1999 07:22 |
| 0041+092 | B05507010 | 17/07/2001 10:29 | 1342+442 | A03404020 | 11/01/2000 23:01 | M10304010 | 20/11/1999 09:02 |
| 0050-332 | M10101010 | 04/07/2000 23:52 | 1440+753 | Z90322010 | 15/01/2003 08:40 | M10306030 | 20/11/1999 10:43 |
|  | P20420010 | 11/12/2000 07:00 | 1528+487 | P20401010 | 27/03/2001 23:14 | M10305030 | 21/11/1999 06:43 |
| 0106-358 | D02301010 | 21/11/2004 06:29 | 1550+130 | E11201010 | 15/04/2004 05:38 | M10305040 | 21/11/1999 10:03 |
| 0131-164 | P20412010 | 10/12/2000 07:45 | 1603+432 | Z90324010 | 02/07/2003 11:51 | M10306020 | 21/11/1999 11:39 |
| 0147+674 | Z90303020 | 12/10/2002 17:08 | 1611-084 | B11904010 | 02/04/2004 11:53 | S30701010 | 14/01/2000 09:40 |
| 0226-615 | A05402010 | 04/07/2000 20:39 | 1615-154 | P20419010 | 29/08/2001 12:40 | M10102020 | 17/02/2000 06:10 |
| 0229-481 | M10504010 | 21/09/2002 11:51 | 1620-391 | Q11001010 | 13/07/2000 14:24 | M10306040 | 09/01/2001 09:02 |
| 0232+035 | P10405040 | 06/01/2004 01:00 | 1620+647 | Z90325010 | 17/05/2002 04:45 | M10305060 | 09/01/2001 09:26 |
| 0235-125 | W56809010 | 30/08/2004 17:27 | 1631+781 | M10528040 | 13/02/2004 20:02 | M10306050 | 10/01/2001 13:20 |
| 0236+498 | Z90306010 | 11/12/2002 16:47 | 1635+529 | C05002010 | 13/07/2002 14:00 | M10305070 | 10/01/2001 13:45 |
| 0252-055 | A05404040 | 28/11/2001 06:39 | 1636+351 | P20402010 | 28/03/2001 14:22 | M10304030 | 10/01/2001 15:08 |
| 0310-688 | U10628010 | 28/05/2006 03:05 | 1648+407 | Z90326010 | 12/07/2002 11:22 | M10306060 | 23/01/2001 06:08 |
|  | D90703010 | 03/11/2003 19:46 | 1711+668 | E90304010 | 09/06/2004 21:30 | M10305080 | 23/01/2001 07:55 |
| 0320-539 | D02302010 | 09/09/2003 08:50 | 1725+586 | Z90327010 | 13/05/2002 23:16 | M10304040 | 23/01/2001 11:18 |
| 0346-011 | B12201020-50 | 20/12/2003 04:43 | 1734+742 | U10384010 | 11/03/2007 08:04 | M10306070 | 25/01/2001 04:46 |
| 0353+284 | B05510010 | 02/01/2001 22:31 | 1800+685 | M10530010-70 | 09/09/2002 15:14 | M10305090 | 25/01/2001 06:33 |
| 0354-368 | B05511010 | 12/08/2001 11:18 |  | P20410010-20 | 01/10/2001 04:55 | M10304050 | 25/01/2001 09:53 |
| 0416+402 | Z90308010 | 05/10/2003 13:39 | 1819+580 | Z90328010 | 11/05/2002 01:25 | M10306080 | 28/09/2001 13:50 |
| 0455-282 | P10411030 | 07/02/2000 09:36 | 1844-223 | P20405010 | 28/04/2001 20:01 | M10305100 | 28/09/2001 15:35 |
| 0457-103 | A05403030 | 20/12/2003 01:23 | 1845+683 | Z90329010 | 08/05/2002 20:40 | M10304060 | 28/09/2001 17:15 |
| 0501+524 | Coadded* |  |  | Z99003010 | 13/09/2002 10:39 | M10306090 | 21/11/2001 09:54 |
| 0512+326 | A05407070 | 01/03/2000 04:08 | 1917+599 | Z90330010 | 07/05/2002 01:36 | M10305110 | 21/11/2001 11:39 |
| 0549+158 | P20417010 | 04/11/2000 15:32 | 1921-566 | A05411110 | 24/05/2000 22:59 | M10304070 | 21/11/2001 13:19 |
| 0603-483 | Z90309010 | 31/12/2002 19:26 | 1942+499 | Z90331010 | 08/05/2002 00:12 | M10306100 | 17/02/2002 07:27 |
| 0621-376 | P10415010 | 06/12/2000 05:28 | 1950-432 | Z90332010 | 04/10/2002 09:18 | M10305120 | 17/02/2002 12:34 |
| 0659+130 | B05509010 | 05/03/2001 21:53 | 2000-561 | Z90333010 | 15/04/2002 21:30 | M10304080 | 17/02/2002 17:43 |
| 0715-704 | M10507010 | 15/08/2003 00:08 | 2004-605 | P20422010 | 21/05/2001 21:09 | M10306110 | 23/02/2002 02:05 |
| 0802+413 | Z90311010 | 14/03/2004 09:37 | 2011+398 | M10531020 | 23/10/2003 12:19 | M10305130 | 23/02/2002 06:43 |
| 0809-728 | U10716010 | 22/01/2006 05:04 | 2014-575 | Z90334010 | 16/04/2002 15:55 | M10304090 | 23/02/2002 08:23 |
|  | Z90312010 | 01/05/2002 05:00 | 2020-425 | Z90335010 | 07/06/2002 09:16 | M10306120 | 25/02/2002 02:17 |
|  | Z90312020 | 15/08/2003 12:51 | 2032+248 | B11905010 | 13/07/2001 02:54 | M10305140 | 25/02/2002 06:59 |
| 0830-535 | Z90313010 | 03/05/2002 13:37 | 2043-635 | Z90337010 | 13/04/2002 09:10 | M10306130 | 03/12/2002 21:00 |
| 0905-724 | U10720010 | 20/01/2006 22:14 | 2111+498 | M10532010 | 27/10/2002 16:44 | M10306140 | 06/12/2002 02:30 |
|  | S60134010 | 10/03/2002 16:12 | 2116+736 | Z90338020 | 28/10/2002 01:26 | M10305150 | 06/12/2002 05:16 |
|  | A054050100 | 04/04/2000 11:18 | 2124-224 | P20406010 | 08/10/2001 20:26 | M10520010 | 07/12/2002 21:46 |
| 0937+505 | Z90314010 | 24/03/2003 09:13 | 2124+191 | A05409090 | 12/07/2001 22:23 | M10306150 | 08/12/2002 22:36 |
| 1019-141 | P20415010 | 15/05/2001 21:35 | 2146-433 | Z90339010 | 06/10/2002 15:14 | M10305160 | 09/12/2002 00:29 |
| 1021+266 | B05508010 | 01/05/2001 19:34 | 2152-548 | M10515010 | 24/09/2002 03:07 | M10304120 | 09/12/2002 03:53 |
| 1024+326 | B05512010 | 01/05/2001 09:22 | 2211-495 | M10303160 | 24/05/2003 23:28 | M10306160 | 05/02/2003 19:14 |
| 1029+537 | B00301010 | 25/03/2001 04:40 | 2257-073 | A05410100 | 28/06/2000 11:05 | M10305170 | 05/02/2003 21:07 |
| 1040+492 | Z00401010 | 04/04/2002 12:42 | 2309+105 | P20424010 | 14/07/2001 01:26 | M10304130 | 06/02/2003 00:36 |
| 1041+580 | Z90317010 | 07/04/2002 10:04 | 2321-549 | Z90342010 | 22/07/2002 05:10 | S40576040 | 21/11/2003 05:21 |
| 1057+719 | Z90318010 | 08/04/2002 16:12 | 2331-475 | M10517010 | 23/09/2002 08:46 | M10306170 | 23/11/2003 20:16 |
| 1109-225 | A05401010 | 29/05/2000 16:00 | 2350-706 | A05408090 | 23/05/2001 16:21 | M10305190 | 25/01/2004 21:31 |
| 1234+481 | M10524020 | 18/03/2003 20:29 |  |  |  | M10304150 | 26/01/2004 00:51 |

## 2.3 Spectral features

Many papers have reported on the spectral features that can be studied in the far-UV wavelength ranges covered by *HST* and *IUE* (~1150 – 1750 Å). A useful summary of the most important/strongest lines can be found in Holberg, Barstow & Sion (1998). For the study of white dwarf photospheres, these include C III (multiplet at 1175 Å), C IV (1548/1550 Å), N V (1238/1242 Å), O V (1371 Å), Si IV (1391/1402 Å) besides numerous features of Fe V/VI and Ni V and formed the basis of the Barstow et al. (2003b) study.

Fig. 1 shows a number of photospheric and interstellar features that are detected in the *FUSE* wavelength range. As in other wavelength ranges for hot DAs, ionization stage is a useful discriminator between interstellar and photospheric species. Typical ISM lines are neutral or singly ionized, while the ionization level of photospheric lines ranges from III to VI. High ionization transitions can be associated with interstellar or circumstellar material. However, as such gas has much lower density than the photosphere, only ground-state resonance features will be seen. In the *FUSE* spectral range, apart from the H Lyman series, only the O VI lines are ground-state transitions. All other high ionization features arise from excited state transitions and must be photospheric. The relative radial velocities of the lines can also be used to determine their origin, as discussed extensively by Barstow et al. (2010), for example, in the context of *FUSE* data. Table 3 gives a comprehensive list of all heavy element features detected in the sample of stars, indicating their likely origin. For measuring photospheric abundances, *FUSE* gives us access to C III, C IV, Si IV, P IV/V and S IV/VI. The usefulness of O VI in the context is compromised by the potential presence of unresolved interstellar or circumstellar contamination, which would render any abundance measurements unreliable. Weak Fe features are detected in a few spectra but not in sufficient numbers to make a detailed study worthwhile. Therefore, in this work we concentrate on measuring the abundance patterns of C, Si, P and S.

Included in this paper are a number of *FUSE* spectra of the metal-rich DA star G191-B2B (WD0501+524), which are considered in the context of our larger white dwarf sample. It is worthwhile noting that these *FUSE* data are also part of a unique *FUSE*/STIS data set on this star (Preval et al. 2013), where the coadded *FUSE* spectrum is combined with coadded STIS E140H and E230H spectra that cover the entire UV from 910 Å to 3150 Å with S/N that exceeds 100 in many bands and in which well over 900 photospheric and ISM features have been identified. Preval et al. (2013) present a far more detailed discussion of G191-B2B than is appropriate here.

**Table 3.** Frequently observed spectral lines.

| Element | Ion | Wavelength (Å) | ISM/Photospheric | Element | Ion | Wavelength (Å) | ISM/Photospheric |
|---|---|---|---|---|---|---|---|
| O | I | 919.660 | ISM | O | VI | 1037.613 | ISM or Phot |
| O | I | 921.857 | ISM | O | I | 1039.231 | ISM |
| O | I | 924.950 | ISM | Ar | I | 1048.220 | ISM |
| O | I | 929.517 | ISM | S | IV | 1062.660 | Phot |
| S | VI | 933.378 | Phot | Si | IV | 1066.620 | Phot |
| O | I | 936.629 | ISM | S | IV | 1072.973 | Phot |
| S | VI | 944.523 | Phot | N | II | 1083.990 | ISM |
| O | I | 948.686 | ISM | P | V | 1117.977 | Phot |
| N | I | 953.415 | ISM | Si | IV | 1122.485 | Phot |
| N | I | 953.655 | ISM | Fe | III | 1124.881 | Phot |
| N | I | 953.970 | ISM | Fe | III | 1126.729 | Phot |
| N | I | 954.104 | ISM | P | V | 1128.008 | Phot |
| N | I | 963.990 | ISM | Si | IV | 1128.330 | Phot |
| N | I | 964.626 | ISM | Fe | III | 1128.724 | Phot |
| N | I | 965.041 | ISM | Fe | III | 1129.191 | Phot |
| O | I | 971.738 | ISM | Fe | III | 1130.402 | Phot |
| P | IV | 1030.515 | Phot | Fe | III | 1131.195 | Phot |
| O | VI | 1031.912 | ISM or Phot | N | I | 1134.165 | ISM |
| P | IV | 1033.112 | Phot | N | I | 1134.415 | ISM |
| P | IV | 1035.516 | Phot | N | I | 1134.980 | ISM |
| C | II | 1036.337 | ISM | Fe | VI | 1152.771 | Phot |
| C | II | 1037.018 | ISM | C | III | Multiplet ~1175 | Phot |

## 3 MODEL STELLAR ATMOSPHERE CALCULATIONS

We have utilized the same grid of homogeneous model stellar atmospheres, calculated using the non-LTE code TLUSTY (see Hubeny & Lanz 1995 and references therein), as in Barstow et al. (2003b). These are based on work reported by Lanz et al. (1996) and Barstow, Hubeny & Holberg (1998, 1999) but extended in temperature range up to 120000 K, to encompass the hottest DA stars, although the hottest object in this particular sample is HD223816, at $T_{\rm eff}$ = 76690 K. To take account of the higher element ionization stages that are likely to be encountered in these objects, new ions of O VI, Fe VII/VIII and Ni VII/VIII were added to the model atoms as well as extending the data for important ions such as C IV to include more energy levels. All the calculations were performed in non-LTE with full line-blanketing. To minimize the time taken for the model calculations, it is advantageous to start with input abundance values close to actual values expected. Therefore, in generating the model grid, heavy element abundances were initially fixed at representative values determined for G191-B2B (He/H = $1.0\times10^{-5}$, C/H = $4.0\times10^{-7}$, N/H = $1.6\times10^{-7}$, O/H = $9.6\times10^{-7}$, Si/H = $3.0\times10^{-7}$, Fe/H = $1.0\times10^{-5}$, Ni/H = $5.0\times10^{-7}$), but taking into account that the C IV lines near 1550 Å have subsequently been resolved into multiple components by STIS (Bruhweiler et al. 2000). Stepping away from these to higher and lower abundances then populated the grid. While there have been significant advances in the computational speed and capability of the code since the work of Barstow et al. (2003b), there have been no changes to the input physics and available atomic data for these models. Hence, the 2003 models remain the state-of-the-art. Furthermore, utilization of the same models facilitates direct comparison between the new results reported here and the earlier work. There are some potential improvements to the models that could be made in future. In particular, Preval et al (2012) have identified a large number of previously unrecognized Fe and Ni features in the very high signal-to-noise 912 - 1750Å spectrum of G191-B2B, indicating that the model atoms used in the TLUSTY calculations are incomplete. A detailed programme of calculations is now underway to generate the energy level data and opacities necessary to extend the Fe and Ni model atoms.

## 4 DATA ANALYSIS

### 4.1 Measurement of photospheric abundances

For the earlier study of the *HST* and *IUE* data we developed a statistical spectral analysis technique to quantify the element abundances and their associated uncertainties. We apply that same technique here, which we summarise briefly. We make use of the programme XSPEC (Shafer et al.1991), which adopts a $\chi^2$ minimization technique to determine the model spectrum giving the best agreement with the data. Any single high-resolution far-UV spectrum has a large number of data points (~12500 for *FUSE*), which is hard for the XSPEC program to handle. Furthermore, some regions of the far-UV spectra contain little useful information, where the data are noisy or devoid of detectable spectral features. Therefore, we split the spectra up into smaller sections, selecting wavelength regions that contain the most useful information around the spectral lines listed in Table 3.

Synthetic spectra were computed for the full *FUSE* wavelength range at each value of $T_{eff}$ and log $g$ in the model grid, at the nominal abundances in the converged TLUSTY models. The spectral grid was then extended, by scaling the abundances of the main species within a particular spectral range using the SYNSPEC spectral synthesis program. Models and data were compared using the XSPEC program, considering one element at a time. All lines of a particular ion were fit simultaneously. When examining a spectral range containing absorption from sources other than the photosphere (mostly ISM), these other lines were excluded from the analysis to avoid potential biases of the photospheric fits. For a given star, $T_{eff}$ and log $g$ were allowed to vary as free parameters within the known 1σ error ranges. To verify that a true minimum $\chi^2$ had been found, the $\chi^2$ space was mapped out by calculating the value of $\chi^2$ over a grid stepping through the values of all the variables in small increments. In addition, uncertainties were computed for the abundances by searching the grid for the $\chi^2$ the parameter values corresponding to a $\delta\chi^2$ probability of 68%, equivalent to 1σ. Examples of fits to the lines of the all the species used to determine the abundances are shown in Figure 2. We have chosen an intermediate signal-to-noise star, WD1021+266, for most of the plots in this illustration, except for the P IV lines (P IV is not detected in WD1021+266), which are from the WD2000-561 spectrum. The error bars in each panel are the observational data and the smooth curve the best-fit model. Any lines not modeled in the plots are for different species, not included in the fit as discussed above.

The results of all the abundance measurements are listed in Table 4 ordered by decreasing stellar effective temperature, for those stars where there is a clear detection of at least one element, numbering 33 objects from our sample. Only single ionization stages are available to measure the abundances of C and Si (C III and Si IV respectively), but we report separate measurements for each ionization stage of P (IV and V). There are potentially two ionization stages for S (IV and VI), but at the short wavelengths of the S VI features, the data are universally too noisy to provide meaningful abundance measurements. Therefore, we only report the S IV abundances. Not all elements are detected in each star. In our total sample of 89 objects we detect no elements other than hydrogen in the remaining 56 cases. We have not determined formal upper limits to the elemental abundances in these, but the lower detection values included in Table 4 give an indication of the threshold sensitivity for the sample: $6.08 \times 10^{-9}$ for carbon, $3.84 \times 10^{-8}$ for silicon, $1.49 \times 10^{-9}$ for phosphorous and $5.92 \times 10^{-9}$ for sulphur.

### 4.2 Analysis of the observed abundances

The measurement of the abundances of C, Si, P and S for 33 stars in a total sample of 89 objects, described above and listed in Table 4, has generated a complex data set that requires careful study and discussion. To facilitate this process we breakdown our analysis of the results into several themes. First, we consider the overall pattern of the detection of heavy elements in the sample as a function of $T_{eff}$. Secondly, we examine how the measured abundances vary across the sample. Finally, in section 5, we consider the relative abundances of the elements detected.

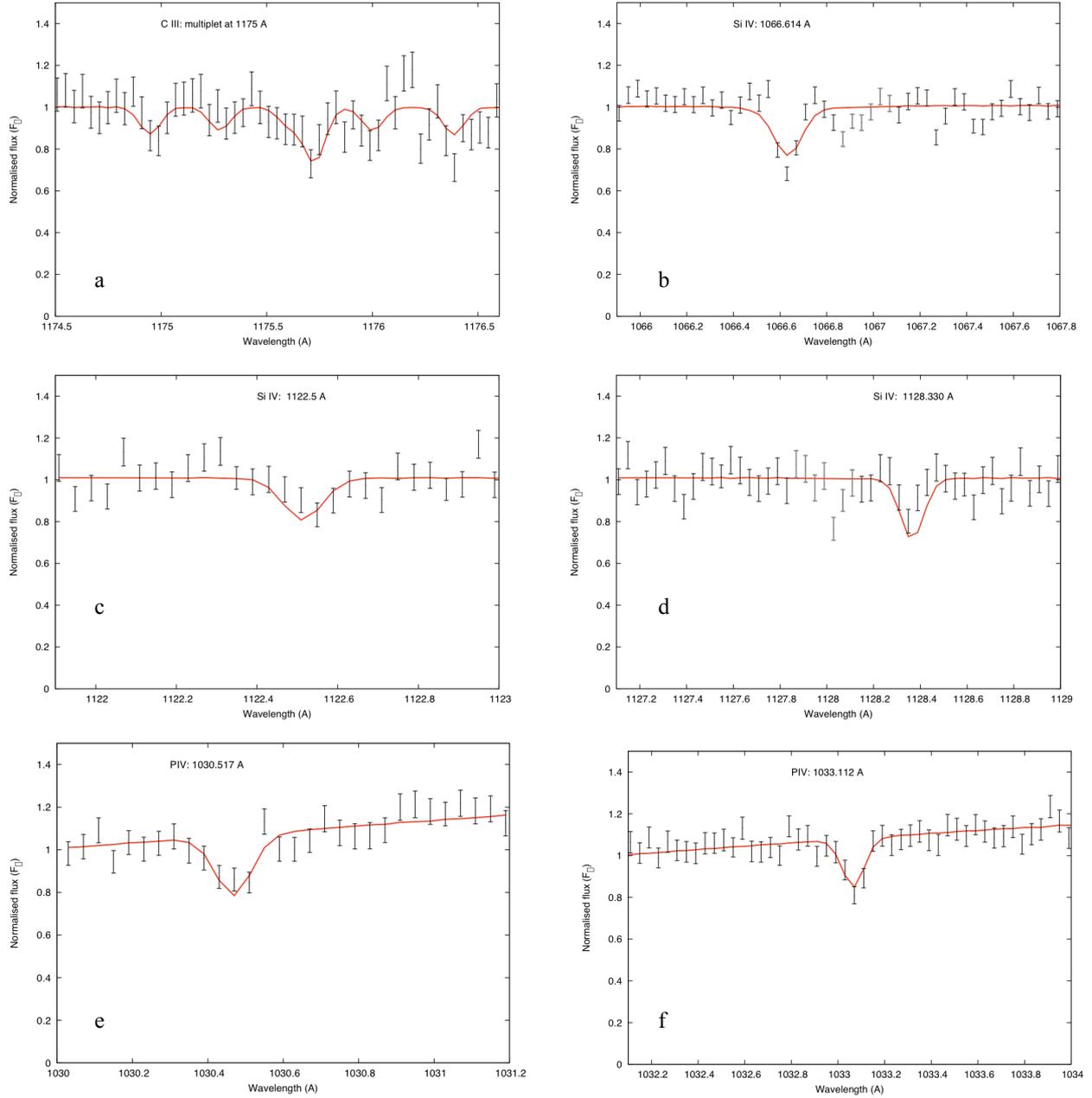

**Figure 2.** Examples of fits to the lines of the all the species used to determine the abundances. We have chosen an intermediate signal-to-noise star, WD1021+266, for most of the plots in this illustration, except for the P IV lines, which are from the WD2000-561 spectrum. The error bars in each panel are the observational data and the smooth curve the best fit to the data. Lighter shaded features in the data are lines that are not included in the fit. a) C III multiplet; b) Si IV 1066.614 Å; c) Si IV 1122.5 Å; d) Si IV 11280.33 Å; e) P IV 1030.517 Å; f) P IV 1033.112 Å.

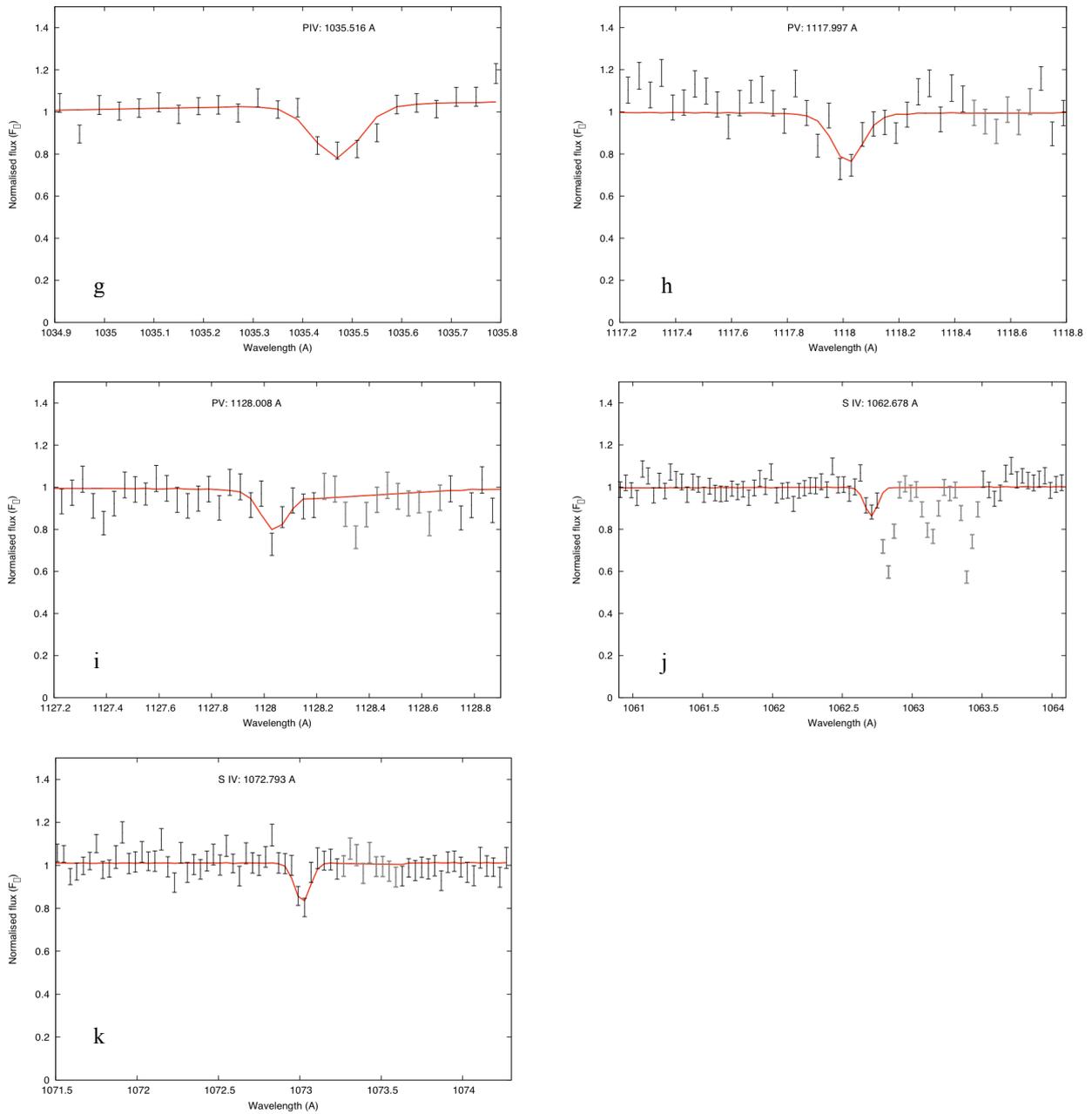

**Figure 2** continued. g) P IV 1035.516 Å; h) P V 1117.997 Å; i) P V 1128.008 Å; j) S IV 1062.678 Å; k) S IV 1072.793 Å.

**Table 4.** Measured abundances for each star, listed in order of decreasing $T_{\text{eff}}$, in the sample for all elements included in this study. Each group of three columns, corresponding to a particular element/ion, lists the actual abundance value followed by the -1σ and +1σ uncertainties respectively. Entries that do not have uncertainties indicate that individual elements were not detected at the limit of the available signal-to-noise of the data. The abundance value listed is then an upper limit. All values are expressed as a number fraction with respect to hydrogen.

| Star | $T_{\text{eff}}$ | C/H | -1σ | +1σ | Si/H | -1σ | +1σ | PIV/H | -1σ | +1σ | PV/H | -1σ | +1σ | SIV/H | -1σ | +1σ |
|---|---|---|---|---|---|---|---|---|---|---|---|---|---|---|---|---|
| WD2350-706 | 76690 | 2.64E-08 | | | 8.18E-07 | 1.43E-07 | 2.08E-07 | 4.53E-08 | 2.85E-08 | 3.25E-08 | 7.51E-08 | 6.80E-09 | 1.83E-08 | 1.74E-07 | 7.80E-08 | 7.85E-08 |
| WD2146-433 | 67912 | 7.04E-08 | | | 3.00E-06 | 2.18E-07 | 9.00E-12 | 1.67E-07 | 8.97E-08 | 8.28E-08 | 9.77E-08 | 2.80E-08 | 5.43E-08 | 3.87E-07 | 1.41E-07 | 2.99E-07 |
| WD1342+442 | 66750 | 5.14E-08 | 2.66E-08 | 4.32E-08 | 2.90E-07 | 1.28E-07 | 3.61E-07 | 5.15E-08 | 4.90E-08 | 1.36E-07 | 3.65E-08 | 1.84E-08 | 3.21E-08 | 5.17E-07 | 3.41E-07 | 7.95E-07 |
| WD2211-495 | 65600 | 2.28E-08 | | | 2.41E-07 | 2.04E-07 | 5.21E-07 | 6.07E-09 | 3.57E-09 | 1.04E-07 | 9.11E-09 | 6.61E-09 | 1.03E-07 | 2.22E-07 | | |
| WD0229-481 | 63400 | 2.66E-09 | | | 8.82E-07 | 1.59E-07 | 4.47E-07 | 6.08E-08 | 2.55E-08 | 4.57E-08 | 5.62E-08 | 1.66E-08 | 1.65E-08 | 1.05E-07 | 4.61E-08 | 9.18E-08 |
| WD0232+035 | 62947 | 7.32E-10 | | | 7.91E-07 | 5.12E-08 | 5.11E-08 | 6.24E-08 | 8.89E-09 | 8.88E-09 | 7.44E-08 | 5.22E-09 | 8.33E-09 | 1.53E-07 | 2.78E-08 | 2.78E-08 |
| WD0621-376 | 62280 | 2.14E-08 | 2.06E-09 | 2.05E-09 | 7.57E-07 | 3.07E-08 | 3.06E-08 | 3.07E-08 | 3.93E-09 | 3.95E-09 | 6.93E-08 | 2.42E-09 | 2.43E-09 | 1.83E-07 | 2.37E-08 | 2.37E-08 |
| WD0455-282 | 58080 | 1.48E-09 | | | 5.12E-07 | 7.22E-08 | 7.21E-08 | 1.80E-08 | 6.87E-09 | 6.85E-09 | 6.08E-08 | 6.95E-08 | 6.92E-08 | 1.52E-08 | | |
| WD0501+524 | 57340 | 1.15E-07 | 3.49E-09 | 3.49E-09 | 5.97E-07 | 1.68E-08 | 1.74E-08 | 2.08E-08 | 1.13E-09 | 1.13E-09 | 2.50E-08 | 1.77E-10 | 5.29E-10 | 6.88E-08 | 3.08E-09 | 3.08E-09 |
| WD2331-475 | 56682 | 2.00E-09 | | | 4.95E-07 | 8.32E-08 | 8.30E-08 | 3.05E-08 | 8.69E-09 | 1.21E-08 | 5.27E-08 | 4.83E-08 | 8.49E-09 | 7.34E-08 | 1.56E-08 | 1.56E-08 |
| WD2309+105 | 51300 | 1.48E-09 | | | 3.84E-08 | 5.23E-09 | 5.27E-09 | 2.15E-09 | 1.15E-09 | 2.05E-09 | 1.16E-08 | 1.60E-09 | 1.70E-09 | 9.78E-09 | 2.79E-09 | 3.75E-09 |
| WD2124-224 | 48297 | 1.68E-09 | | | 7.46E-08 | 2.08E-08 | 2.07E-08 | 1.22E-09 | | | 6.55E-10 | | | 1.68E-08 | | |
| WD2011+398 | 47057 | 3.71E-07 | 3.98E-08 | 2.90E-08 | 7.42E-08 | 1.30E-08 | 1.31E-08 | 2.13E-09 | 1.13E-09 | 3.67E-09 | 2.76E-09 | 6.21E-10 | | 1.86E-06 | 4.63E-08 | 1.37E-08 | 1.34E-08 |
| WD0001+433 | 46205 | 1.48E-08 | | | 8.90E-08 | 6.00E-08 | 7.30E-08 | 2.49E-08 | 2.39E-08 | 7.51E-08 | 2.20E-09 | | | 2.16E-08 | | |
| WD2321-549 | 45860 | 6.08E-09 | 2.08E-08 | 4.84E-08 | 1.39E-07 | 4.36E-08 | 5.14E-08 | 8.92E-08 | 6.05E-09 | 1.02E-08 | 6.71E-09 | 2.56E-09 | 2.57E-09 | 1.04E-08 | 6.36E-09 | 1.09E-08 |
| WD1819+580 | 45330 | 1.34E-07 | 2.49E-08 | 6.09E-08 | 6.84E-07 | 1.32E-07 | 1.48E-07 | 7.68E-09 | 6.44E-09 | 1.13E-08 | 5.07E-09 | 2.32E-09 | 2.73E-09 | 4.44E-09 | | |
| WD1029+537 | 44980 | 2.17E-07 | 8.56E-08 | 8.32E-08 | 8.77E-08 | 2.46E-08 | 4.63E-08 | 3.50E-09 | | | 9.36E-09 | 3.11E-09 | 5.74E-09 | 1.22E-08 | 6.37E-09 | 9.17E-09 |
| WD0131-164 | 44850 | 2.08E-09 | | | 6.97E-08 | 1.65E-08 | 1.65E-08 | 3.30E-09 | | | 9.21E-09 | 2.26E-09 | 4.51E-09 | 2.20E-09 | | |
| WD2000-561 | 44456 | 1.91E-08 | 8.29E-09 | 8.95E-09 | 6.77E-08 | 2.65E-08 | 2.65E-08 | 5.59E-08 | 2.83E-08 | 3.45E-08 | 2.45E-08 | 7.48E-09 | 1.25E-08 | 3.18E-08 | 1.53E-08 | 2.43E-08 |
| WD0512+326 | 42849 | 3.48E-09 | | | 1.54E-07 | 4.87E-08 | 4.87E-08 | 1.00E-07 | 9.90E-08 | 6.00E-13 | 9.10E-11 | | | 6.20E-08 | 3.95E-08 | 5.65E-08 |
| WD1950-432 | 41339 | 6.88E-09 | 2.88E-09 | 3.96E-09 | 1.53E-07 | 4.81E-08 | 4.79E-08 | 3.90E-09 | | | 2.35E-10 | | | 2.13E-09 | | |
| WD2111+498 | 38866 | 1.40E-08 | | | 8.87E-07 | 7.99E-08 | 8.01E-08 | 4.48E-09 | 3.48E-09 | 6.52E-09 | 5.46E-08 | 9.49E-09 | 9.41E-09 | 5.92E-09 | 2.92E-09 | 4.43E-09 |
| WD1611-084 | 38500 | 4.00E-07 | 1.04E-08 | 0.00E+00 | 5.99E-08 | 1.96E-08 | 1.96E-08 | 3.50E-09 | | | 4.26E-09 | 3.25E-09 | 3.84E-09 | 1.95E-09 | | |
| WD1021+266 | 35432 | 7.77E-09 | 3.77E-08 | 6.23E-08 | 7.51E-08 | 2.81E-08 | 3.04E-08 | 1.49E-09 | 4.88E-10 | 8.01E-09 | 1.74E-08 | 1.03E-08 | 1.46E-08 | 1.55E-08 | 6.16E-09 | 9.38E-09 |
| WD0050-332 | 34684 | 2.84E-10 | | | 1.96E-09 | | | 2.10E-09 | | | 8.68E-09 | 1.37E-09 | 1.38E-08 | 5.52E-10 | | |
| WD1942+499 | 33500 | 8.80E-10 | | | 1.74E-07 | 3.04E-08 | 3.03E-08 | 1.85E-09 | 8.48E-10 | 1.16E-08 | 3.83E-08 | 1.01E-08 | 1.40E-08 | 3.84E-09 | | |
| WD1917+599 | 33000 | 1.76E-09 | | | 1.31E-07 | 4.39E-08 | 6.01E-08 | 1.00E-07 | 8.10E-10 | 1.00E-12 | 2.33E-09 | 1.33E-09 | 3.42E-09 | 3.60E-09 | | |
| WD0549+158 | 32780 | 2.04E-09 | | | 1.44E-08 | 4.40E-09 | 1.01E-08 | 2.75E-09 | | | 3.84E-09 | 1.59E-09 | 2.40E-09 | 8.10E-10 | | |
| WD0252-055 | 30355 | 1.63E-09 | | | 3.62E-08 | 2.52E-08 | 3.68E-08 | 1.29E-08 | 1.19E-08 | 5.65E-08 | 6.49E-08 | 5.03E-09 | 8.36E-09 | 3.42E-09 | | |
| WD1019-141 | 31340 | 1.07E-08 | 2.70E-09 | 4.18E-09 | 7.82E-08 | 2.82E-08 | 3.28E-08 | 5.95E-08 | 4.92E-08 | 4.05E-08 | 2.60E-09 | | | 2.67E-09 | | |
| WD1734+742 | 28795 | 3.64E-08 | 1.31E-08 | 2.68E-08 | 4.67E-07 | 1.98E-07 | 2.57E-07 | 9.88E-08 | | | 5.80E-09 | | | 1.05E-08 | | |
| WD0106-358 | 28580 | 7.38E-08 | 1.54E-08 | 1.52E-08 | 7.89E-08 | 1.23E-08 | 1.23E-08 | 4.20E-09 | | | 3.76E-08 | 1.23E-08 | 1.97E-08 | 7.14E-09 | | |
| WD1337+701 | 20435 | 2.74E-07 | 7.85E-08 | 7.71E-08 | 7.82E-07 | 1.33E-07 | 1.31E-07 | 9.98E-08 | 9.88E-08 | 2.35E-10 | 1.00E-07 | 9.90E-08 | 0.00E+00 | 2.64E-08 | | |

### 4.3 Observed patterns of element abundances

Our sample of 89 stars represents the most comprehensive and self-consistent dataset available for studying how the atmospheric compositions of hot DA white dwarfs evolve, varying with age and mass. For the purposes of this discussion we use the parameters we can measure directly, $T_{\text{eff}}$ and log $g$, as proxies for these. Furthermore, the even coverage of the temperature range allows us to examine the range of compositions observed at any particular temperature.

#### 4.3.1 Detections and non-detections of heavy elements

Table 4 includes about a third of the stars in the sample studied (33 objects), which all have detectable quantities of elements heavier than He in their atmospheres. However, it does not include the remaining 56 stars, where no such material is detected. Consequently, Table 4 does not give an indication of the distribution. For most stars, the detections of heavy elements in the *FUSE* wavelength range are matched by detections in *HST* and/or *IUE*, where these instruments have observed the stars (see Barstow et al. 2003b). Not all elements are detected in all the stars that exhibit some heavy elements and there are two objects, WD2152-548 and WD2032+248, where detections are seen in the *STIS/IUE* spectra but there are no detections in the *FUSE* wavelength range. Since these detections are all resonance lines, it is likely that the material is circumstellar rather than photospheric. Figure 3 shows the fraction of stars with and without detections of any element as a function of effective temperature, with the breakdown for individual elements in Figure 4. Most heavy element detections are seen in at the highest temperatures, with a monotonic decrease in the fraction towards lower temperatures, except for carbon, which has about half the number of detections of the other elements and a relatively flat trend. Figure 5 shows the same information as figure 3, but as a function of cooling age.

#### 4.3.2 Measured element abundances

The abundance of carbon has been measured from the C III multiplet near 1176 Å, where detected. The values are listed in Table 3 along with their associated measurement errors. Out of the whole sample of 89 objects, only 15 stars contain detectable abundances of carbon. Figure 6 shows the carbon abundance as a function of stellar effective temperature along with the values predicted by the radiative levitation calculations of Chayer et al. (1995), where the light and heavy solid curves correspond to log $g$ values of 7.5 and 8.0 respectively, the range occupied by the majority of stars in the sample. Most of the detections are confined to $T_{\text{eff}}$ below ~45000 K. However, three of the hottest stars also have the C III multiplet in their spectra at temperatures where the carbon is mostly ionized to C IV and the C III levels have very low populations, usually rendering the C III lines undetectable. There is little agreement between the observed carbon abundances and those predicted by the radiative levitation calculations, which are generally at least an order of magnitude greater than the former, except in a few cases at the lower end of the temperature range. The abundance detected in the coolest star in this sub-sample, WD1337+706, is the only example in which the C abundance is greater than predicted, by about two orders of magnitude. No particular trend for varying abundance with $T_{eff}$ can be seen. Indeed the spread of abundances at a particular temperature seems to be uniform across the temperature range, within the limited statistics.

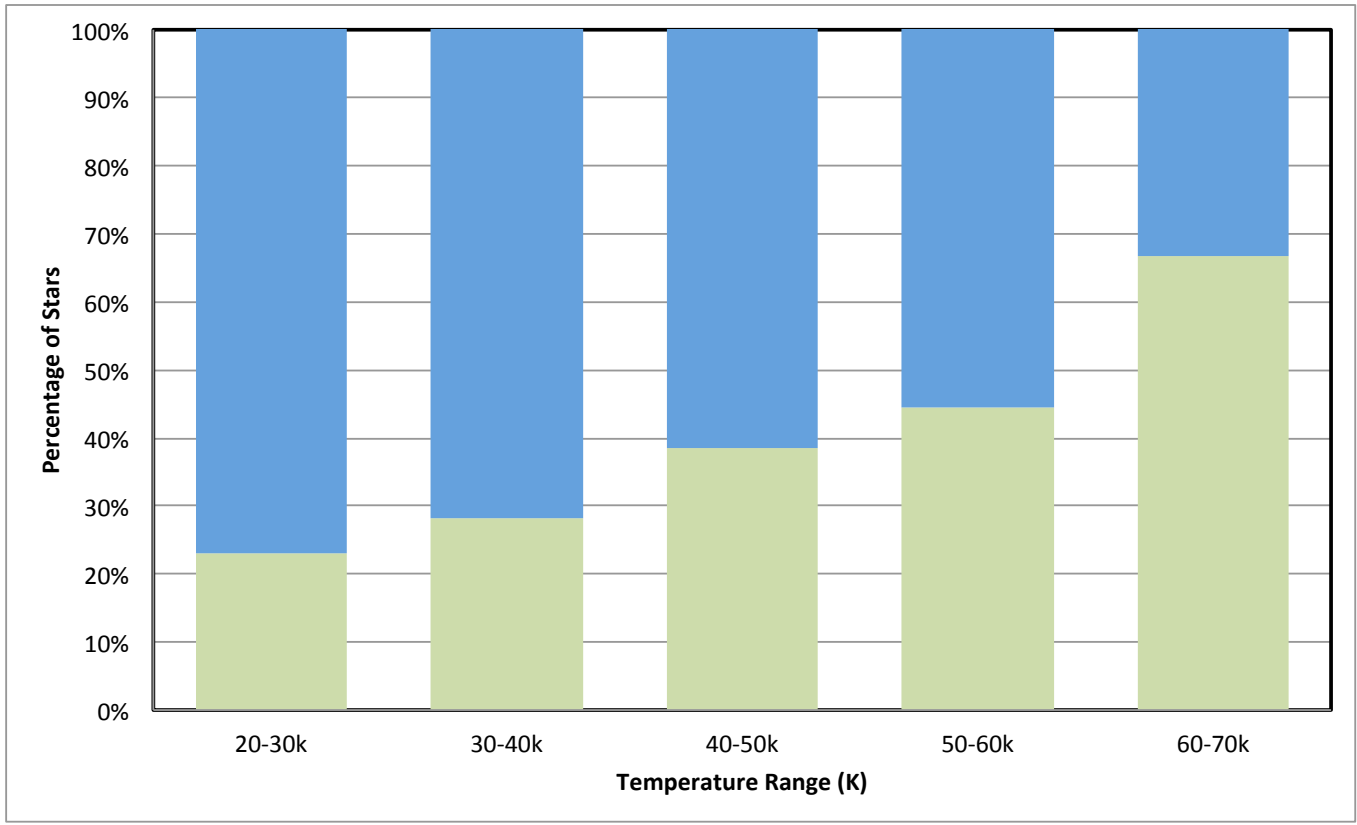

**Figure 3.** The fraction of stars with detected heavy elements (green) and without detected heavy elements (blue) as a function of effective temperature, in 10000 K bins.

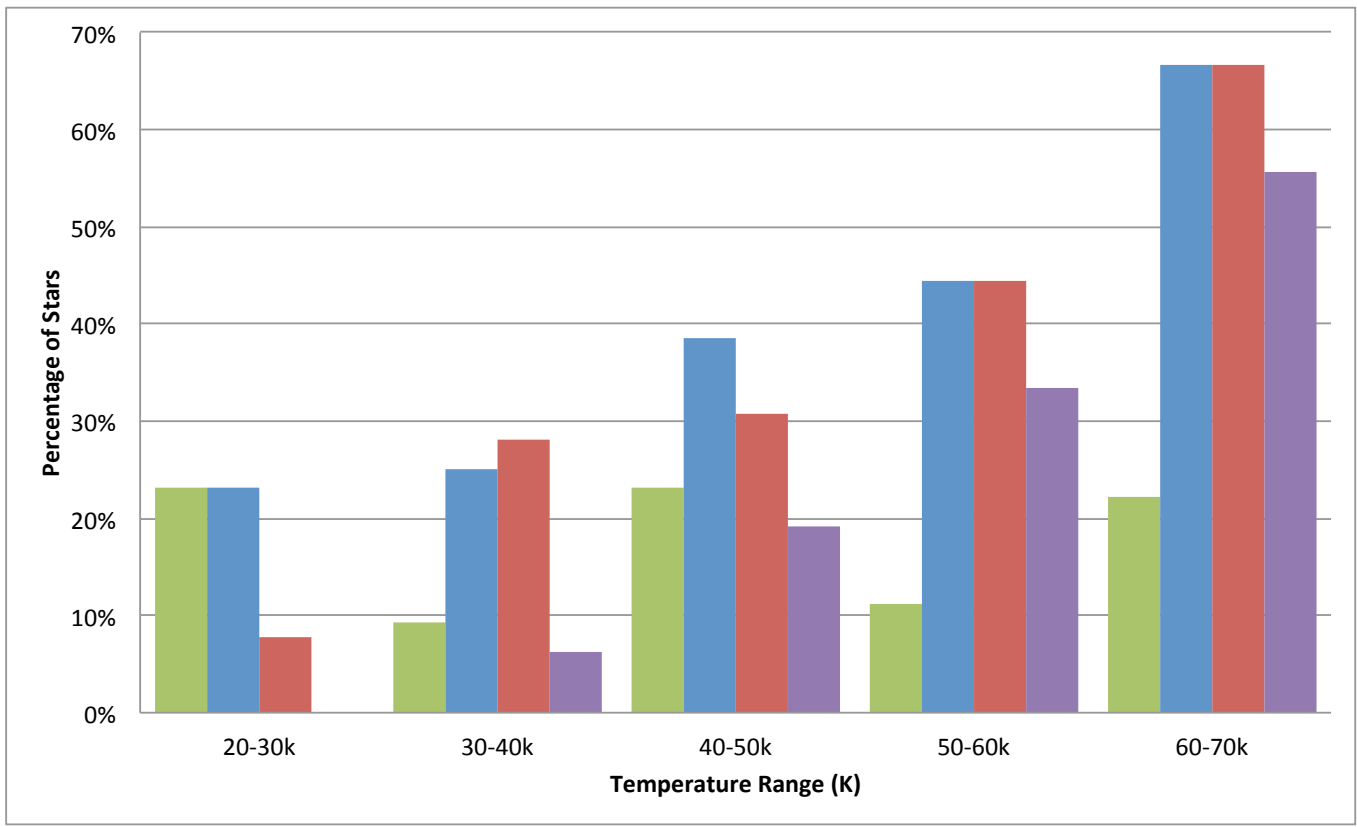

**Figure 4.** The fraction of stars with detected heavy elements as a function of effective temperature, in 10000 K bins by element: carbon (green), silicon (blue), phosphorous (red) and sulphur (purple).

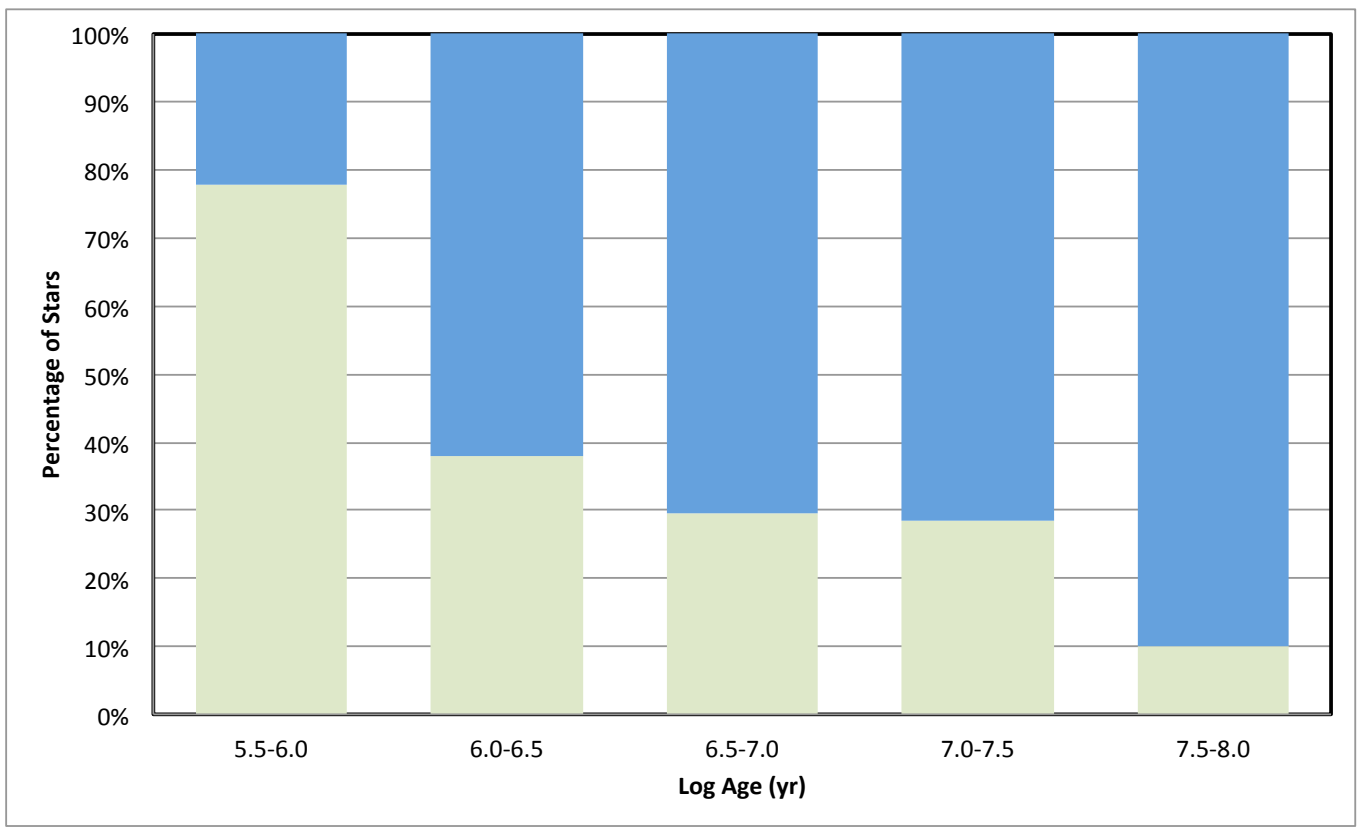

**Figure 5.** The fraction of stars with detected heavy elements (green) and without detected heavy elements (blue) as a function of log cooling age.

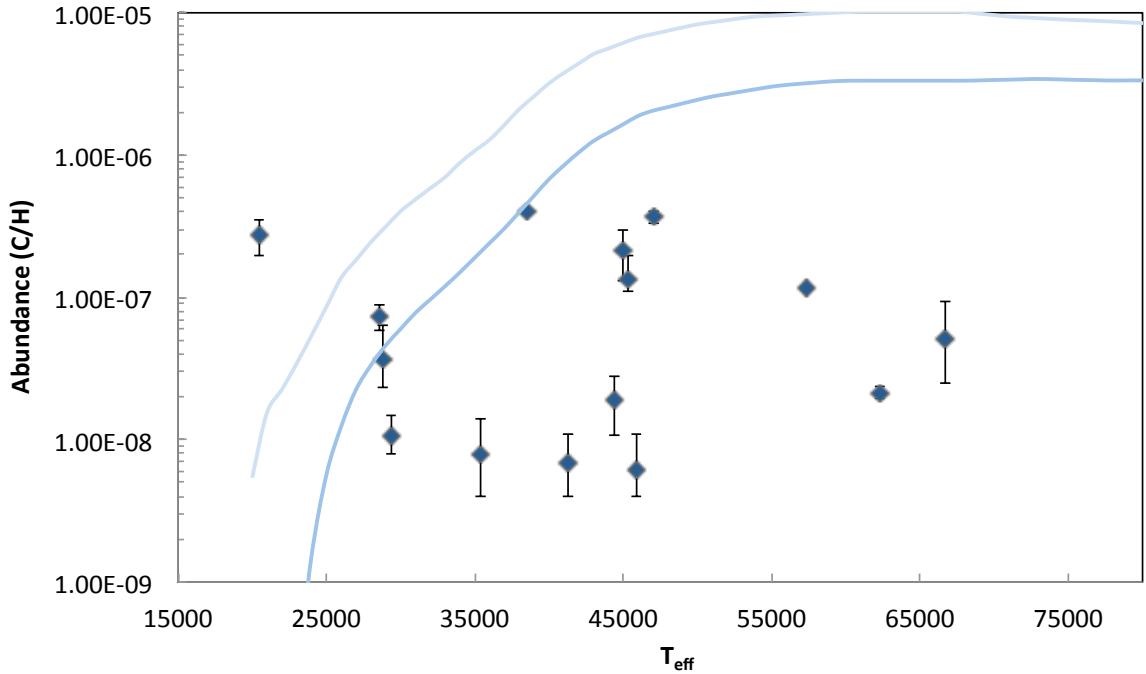

**Figure 6.** Measured abundances of carbon (by number with respect to hydrogen, filled diamonds) as a function of $T_{eff}$. The solid curves are the values predicted for carbon by the radiative levitation calculations of Chayer et al. (1995) for log $g$ of 7.5 (light curve) and log $g$ of 8.0 (heavy curve), the range encompassing most the objects in the sample.

The most frequently detected element in these stars is silicon, found in 32 objects (see Table 3 and Figure 7). Figure 7 shows the measured abundances of Si as a function of $T_{eff}$ and the radiative levitation predictions. The detections are more evenly spread over the sample temperature range than those of carbon and in a few cases the measured abundances lie in the range predicted. However, for most objects the predicted abundances are inconsistent with the observations. Between 25000 K and 45000 K, a number of stars have lower abundances than expected while, above 55000 K, where the Si abundance is expected to be negligible there are stars with significant abundances. There is also a weak trend of increasing abundance with increased effective temperature. Below 55000 K there is no correlation between abundance and temperature.

Abundance measurements of phosphorus are available from both P IV and P V lines. The results are displayed separately in Figures 8 and 9. There is a large degree of overlap in the objects included in both Figures, with P IV and P V usually detected together, but there are a few objects where only one species is present. The measurement errors on the P IV values are typically larger than those for P V, probably due to a combination of differences in signal-to-noise in the particular spectral region, which is better at the longer wavelengths covered by P V, and sensitivity of the line strength to different abundances. However, within the overall measurement uncertainties, there is good agreement between the abundances measured using the difference ionization stages. The overall pattern of P abundances in Figures 8 and 9 is very similar to that of Si in Figure 7, although the magnitude is about a factor of 10 lower. As for Si, there is a trend of increasing abundance with increasing temperature above 55000 K. Below this temperature there is considerable scatter in the measured abundances, with the largest values being similar to those in the high temperature group. Phosphorus is mostly only detected when silicon is also present. There is clearly no match between observed abundances and the radiative levitation predictions. A few points do agree, but this is most likely coincidence, where the curves cross the range of observed values near 30000-35000 K.

Figure 10 shows the abundance of sulphur as a function of temperature. Here there is a clear correlation of increased abundance with effective temperature rising from a minimum at $T_{eff} \sim 35000$ K. There is also an increase in abundance to lower temperatures. There are fewer detections of sulphur than of phosphorus and, apart from one case (WD1056+516), it is only present when phosphorous is detected.

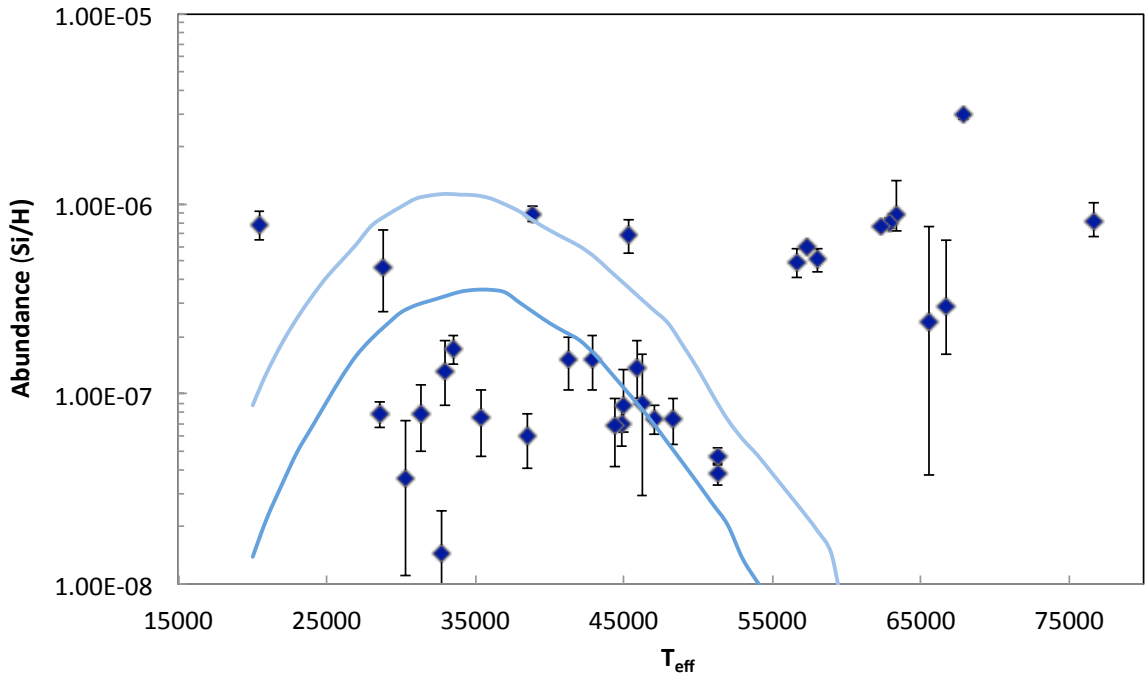

**Figure 7.** Measured abundances of silicon (by number with respect to hydrogen, filled diamonds) as a function of $T_{eff}$. The solid curves are the values predicted for silicon by the radiative levitation calculations as in Figure 6.

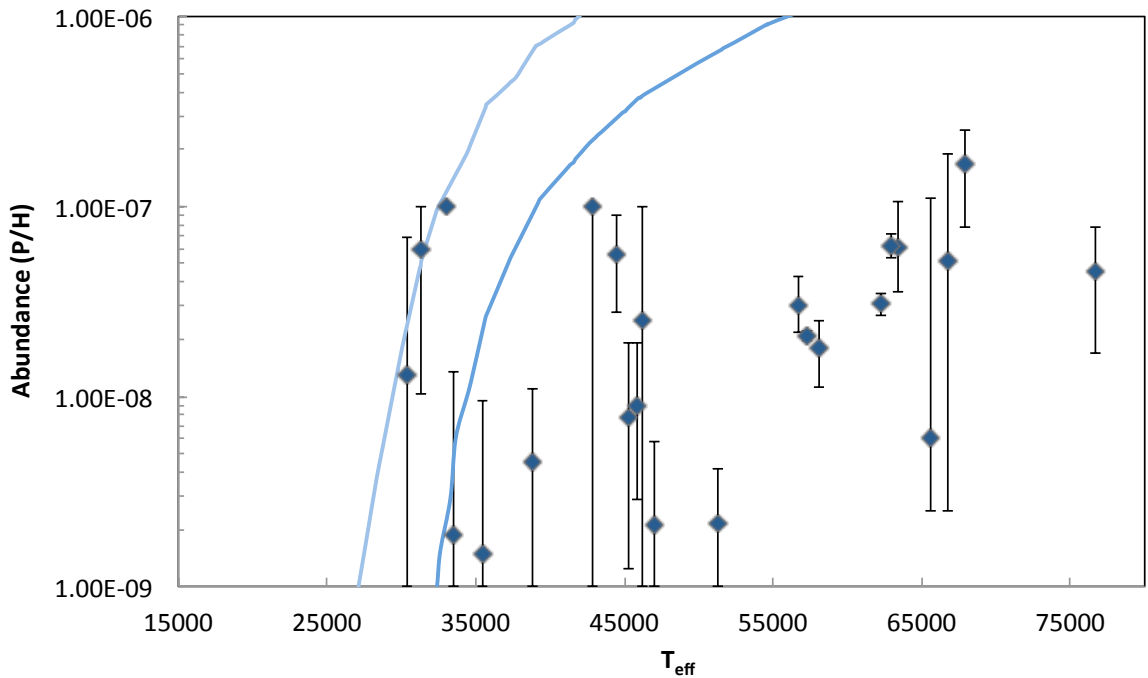

**Figure 8.** Measured abundances of phosphorus (by number with respect to hydrogen, filled diamonds) as a function of $T_{eff}$, from detections of P IV. The solid curves are the values predicted for phosphorus by the radiative levitation calculations of Vennes et al. (1996) for log $g$ of 7.5 (light curve) and log $g$ of 8.0 (heavy curve), the range encompassing most of the objects in the sample.

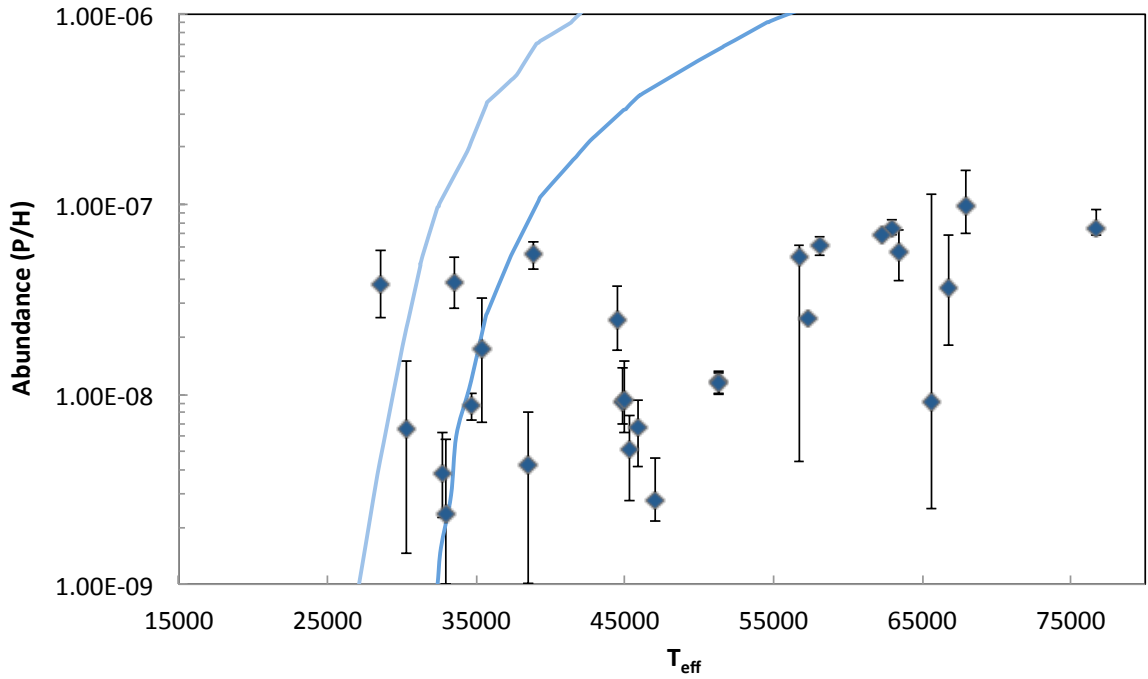

**Figure 9.** Measured abundances of phosphorus (by number with respect to hydrogen, filled diamonds) as a function of $T_{\text{eff}}$, from detections of P V. The solid curves are the values predicted by the radiative levitation calculations as in Figure 8.

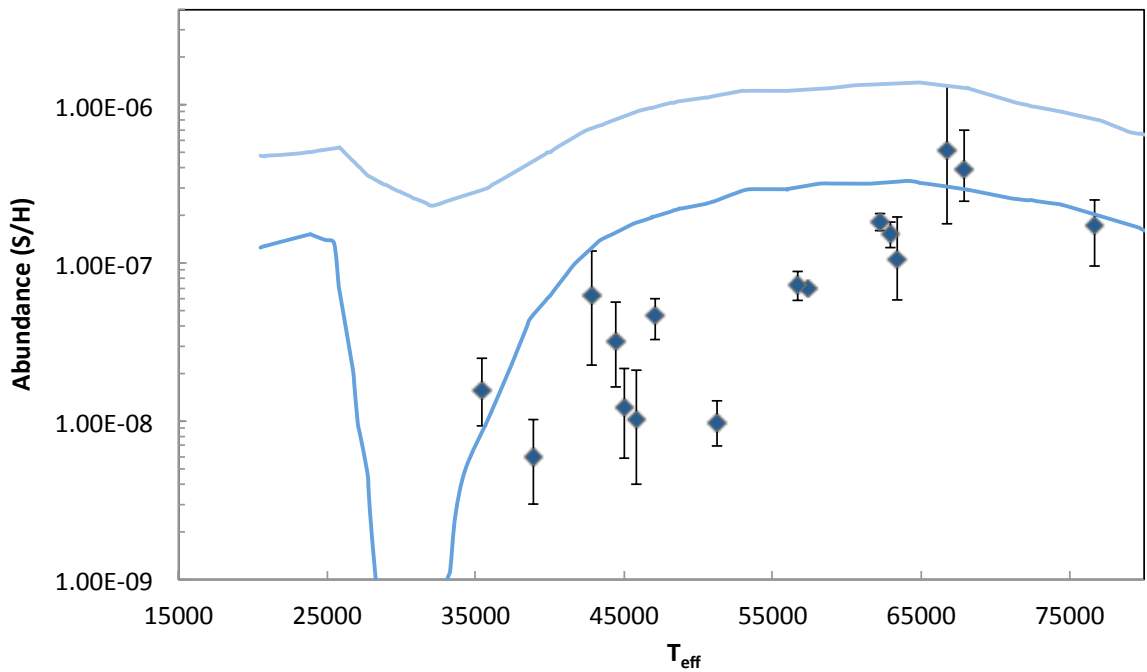

**Figure 10.** Measured abundances of sulphur (by number with respect to hydrogen, filled diamonds) as a function of $T_{\text{eff}}$, from detections of S IV.

# 5 DISCUSSION

As in our earlier work a key element of this study has been the careful objective approach taken to measure abundances with all available lines of a particular species treated simultaneously, using a $\chi^2$ goodness of fit, as described in section 4. This provides a firm statistical basis for determining the measurement uncertainties on which much of the analysis and the discussion in this section is based. Importantly, it allows us to treat correctly the non-uniformity of the signal-to-noise in the observations, which vary considerably across the sample due to differences in relative brightness and exposure times. Sometimes, as can be seen in Figures 4-8, the sizes of the uncertainties in the abundance measurements are large compared to the magnitude and range of abundances observed. However, in general the uncertainties are sufficiently small to allow some clear conclusions to be drawn, some of which contradict the results from the Barstow et al. (2003b) sample. We list these below and then discuss them in more detail in the subsequent subsections.

1. Stars that have atmospheres containing heavy elements and stars with apparently pure H envelopes are present across the whole temperature range. This is not what would be expected if composition were determined entirely by radiative levitation.
2. Where heavy elements are present, the measured abundances are not well matched by the predicted values.
3. Few strong temperature/evolutionary effects are seen in the abundance measurements for the hottest objects.

## 5.1 Radiative levitation and element abundances

The role of radiative levitation as a mechanism for supporting heavy elements in a white dwarf atmosphere against the downward force of gravity has been well established, as discussed in key papers by Chayer Fontaine & Wesemael (1995) and Chayer et al. (1994, 1995). The general pattern of EUV opacity with stellar temperature and gravity, mapped out by photometric observations, provided strong supporting evidence (e.g. Barstow et al. 1993; Marsh et al. 1997; Vennes et al. 1996). Conversely, the predicted abundances have always been in conflict with measured values (e.g. Holberg et al., 1993, 1994; Chayer, Fontaine & Wesemael 1995; Chayer et al. 1994, 1995). The work of Barstow et al. (2003b) provided further supporting evidence that radiative levitation was the key process. However, the enlarged sample included in this work contradicts this view substantially.

In considering the distribution of heavy elements with temperature shown in Figures 6 to 10 it is evident that, with the possible exception of Figure 8 containing the S abundances from S IV, the observed abundances exhibit none of the characteristics of the theoretical equilibrium abundances. In fact the observed distributions demonstrate little regard for effective temperature, but rather occur in broad bands of abundance over some 2 orders of magnitude from the hottest to the coolest stars in our sample. This situation is particularly evident in Figure 9, which shows phosphorus abundance with temperature. In effect, equilibrium abundances are very poor predictors of observed abundances. This finding is similar to what was demonstrated by Barstow et al. (2003b) at longer wavelengths for a somewhat different set of ions. Explanations are required: (1) why do stars of similar temperatures and gravities exhibit very different abundances, (2) why do some stars have abundances very much larger than predicted (see for example Figure 7) and (3) why do certain stars have very low, or immeasurably low, abundances with respect to predictions (see Figures 8 and 9 for example)? It does not seem possible to accommodate all of these observations within the scope of simple radiative equilibrium calculations.

The likely deficiencies in the theoretical framework of Chayer et al. have been acknowledged in their papers. For example, the equilibrium calculations were not carried out self-consistently, with the effects of the element distributions folded into a recalculation of the atmospheric structure. Dreizler & Wolff (1999) and Schuh et al. (2002) made improvements in the approach by incorporating the radiative levitation calculations self-consistently into their atmosphere code. In particular, Schuh et al. were successful in matching the observed *EUVE* spectra of a sample of stars with synthetic spectra, using just $T_{eff}$ and $\log g$ as the free parameters, from which the element abundances are determined within the models. From this work they derive a metallicity parameter, which showed a general decrease with decreasing temperature. However, the values of $T_{eff}$ and $\log g$ determined from their fits were not consistent with those independently measured from the Balmer lines. Secondly, stars that have quite different abundances measured using the UV features were found to have similar metallicity from the EUV analyses.

A particular assumption made in the radiative levitation calculations of all authors is that there is a reservoir of heavy elements that is sufficient to supply the equilibrium abundances. However, this has hardly been examined in detail. The extreme contrast between stars that have apparently pure H and metal containing atmospheres cannot be explained by the modest differences in temperature and gravity alone. Irrespective of any perceived limitations of the radiative levitation calculations, stars with similar physical properties should have similar atmospheric compositions unless their sources of heavy elements are different in some way.

## 5.2 Potential sources of heavy elements

We are studying field white dwarfs, a population with a mean mass close to 0.6 $M_\odot$ and very narrow mass distribution (see e.g Kleinmann et al. 2013). Studies of the relationship between the initial progenitor mass and final white dwarf mass, indicate that these stars evolve from objects with masses around 2 (±0.5) M$_\odot$ (e.g. Kalirai et al. 2008; Rubin et al. 2008; Casewell et al. 2009; Dobbie et al. 2009), typical of early F and A stars. All the white dwarfs in this sample are relatively young, ranging from 4 x 10$^5$ to 1.8 x 10$^6$ years, depending on their temperature and gravity (high gravity, smaller radius stars cool more slowly). This timescale is short compared to the typical main sequence lifetimes of around 2 billion years. Therefore, the progenitors to our field white dwarfs will most likely be Population I objects with heavy element content typical of stars like the Sun. An open question is how the heavy elements are then manifested in the white dwarf.

With the present data set we now have a much clearer picture of heavy element abundance patterns in DA stars over a wide range of temperatures. However, with this clearer picture comes the realization that few patterns have become evident and that definitive conclusions seem hard to demonstrate from the observational data. Nevertheless this is certainly telling us something.

If all or most DA stars evolve from the planetary nebulae central star precursors with photospheres containing similar residual amounts of heavy elements then as they cool, radiative levitation would seem to predict declining abundances during cooling. One might claim to see a hint of this in Figure 10, for the S IV ion but in light of the abundance patterns for other ions and elements, this is not convincing. It would appear from our observations that by the time DAs have reached a temperature of approximately 70000 K, they can exhibit a wide range of heavy element abundances and that this situation persists as they cool to lower temperatures. A case can be made from Figures 3 and 5 that this expectation is evident from the declining fraction of stars with heavy elements as effective temperatures diminish, as the stars age. One scenario is that DA white dwarfs are born with very different initial levels of heavy element content. Some may be born with virtually pure H photospheres and continue to remain so as they cool, there being nothing to levitate. Other WDs are born with relatively high levels of heavy elements approaching 10$^{-6}$ for certain ions. For these stars abundances may well diminish but apparently not under the sole control of radiative levitation, since initial overabundances should relax to steady state equilibrium levels and to create sub-photospheric reservoirs that can maintain diffusive equilibrium abundances. As considered in Barstow et al. (2003b) we really do not know much about such hypothetical reservoirs and the roles they play. A second scenario is that there are processes that effectively compete with simple diffusive equilibrium, such as selective mass loss of certain ions and the possibility of external accretion. Selective mass loss has been discussed in terms of the existence of non-photospheric 'circumstellar' resonance lines seen in some hot DA and DO stars. Now, however, such features are perhaps best explained as well-detached fossil Stromgren spheres (see Dickinson et al. 2012 for a discussion of this point).

In the past, accretion from the ISM has been invoked as a possible explanation for the metals seen in white dwarfs (e.g. Dupuis, Fontaine & Wesemael 1993). However, the typical dearth of hydrogen relative to calcium in the DZ stars (see e.g. Dufour et al. 2007) challenges this scenario. Furthermore, a study of the DZ and DC white dwarfs in the Sloan Digital Sky Survey demonstrates that pollution by the ISM cannot simultaneously account for both the metal-polluted and unpolluted subpopulations (Farihi et al. 2010a). An alternative possible source of external accretion derives from debris disks, formed during the white dwarf phase, that are remnants of planetary, asteroidal or cometary material. It is now well established from IR excesses that such disks do exist surrounding some cooler DBZ and DAZ stars (e.g. Farihi et al. 2010b; Melis et al. 2010). The overall incidence of such well-established debris disks is relatively small, about 3 percent of all DA stars above 10000K. However, the incidence of DAZ stars is much higher at about 20% of all DA stars in the same temperature range. Therefore, it is not difficult to imagine that, in the entire population of DAs, there exists a larger fraction of stars with debris but having no clear IR signatures. Indeed, as discussed in the introduction, at white dwarf temperatures above 24000-35000K dust grains will be sublimated to form gaseous material (von Hippel et al. 2007).

One interesting consideration is that asteroidal or planetary debris disks have apparently terrestrial-like patterns of refractory elements (Si, Mg, Fe, Ca etc. e.g. Klein et al. 2011; Jura et al. 2012; Gäensicke et al. 2012). With *FUSE* we now have two elements to consider, S and P, which are much more volatile, in addition to C, which is also detectable in the *IUE* and *HST*/STIS wavelength ranges. While each element is not detected in all stars with metals, there are a sufficient number with detections of more than one of these elements to allow a study of their relative abundances. Figure 11 shows the ratio of Si and C abundances (by number) as a function of $T_{\mathrm{eff}}$, comparing the observational data with the values predicted by the radiative levitation calculations and values for Solar composition, CI chondrite and bulk Earth material. The Si/C ratio is well differentiated in these three different situations. There is no obvious pattern of variation in the C/Si ratio and a strong predicted correlation with temperature is not seen, although in a few cases the observed ratio is close to that of the radiative calculations. There is scatter in the measured values across a range of nearly 3 orders of magnitude but the majority of stars exhibit Si/C ratios that lie between the CI chondrites and bulk Earth, with C strongly depleted compared to Solar values. Figure 12 shows the Si/P ratio. Again there is large scatter but with a weak trend for increased Si/P with temperature. It is striking that; in general, the Si/P ratio is well below even the Solar value, implying strong enrichment of phosphorus compared to Si. This is not the case for the Si/S ratio (Figure 13) where the ratio lies between the Solar/chondrite and bulk Earth values for most stars.

Overall, there is strong evidence that, where heavy elements are detected, the hot DA white dwarfs are accreting extra-solar planetary material, which has a rocky composition. While the cooler stars below temperatures around 20000K are accreting from dusty debris disks, the hotter stars are probably accreting gaseous material. The timescales for these processes are rather different. Rings consisting entirely of solids evolve enormously slowly (Farihi, Zuckerman & Becklin 2008; Metzger, Rafikov & Bochkarev 2012). However, completely gaseous disks should accrete many orders of magnitude faster, on timescales of days to tens of years depending on the largest radial extent (e.g Jura 2008; Farihi et al. 2012). Most of the stars in the sample studied here are likely to be accreting from gas disks, which may be very short-lived.

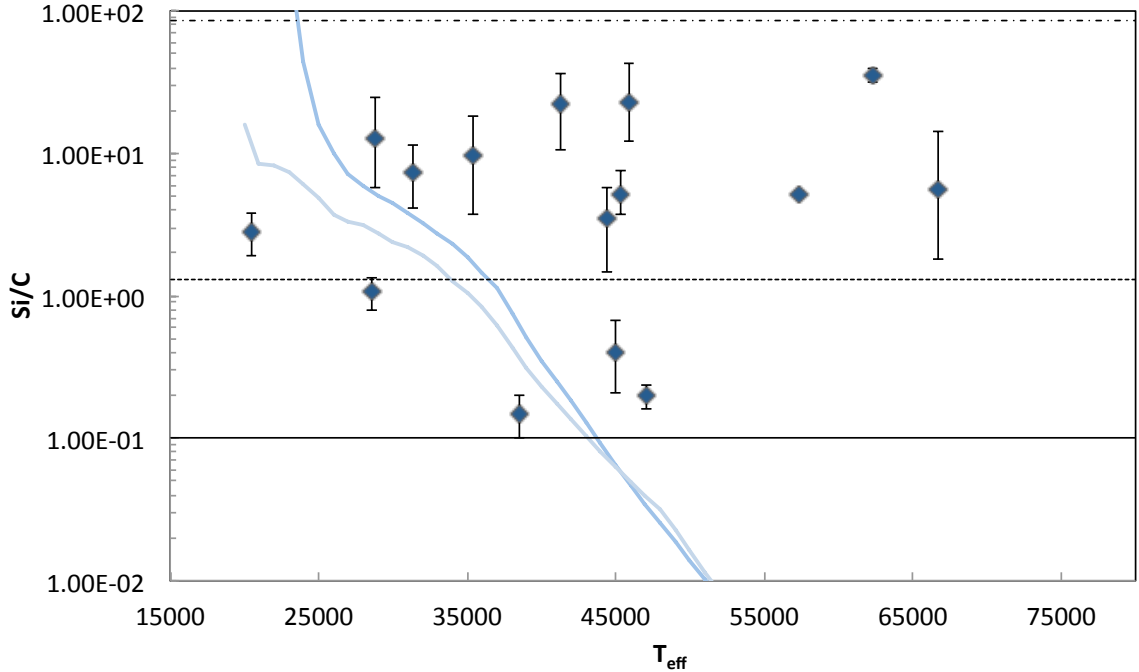

**Figure 11.** The ratio of Si:C abundance (by number) for the white dwarfs where both elements are detected, as a function of $T_{\rm eff}$. The solid curves are the values predicted for the ratio by the radiative levitation calculations of Chayer et al. (1995) for log $g$ of 7.5 (light curve) and log $g$ of 8.0 (heavy curve). The horizontal lines correspond to Solar (solid line, Anders & Grevesse 1989; Grevesse & Noels 1993), CI chondrite (dashed line, Lodders 2003) and bulk Earth (dash-dot line, Allègre, Manhès & Lewin 2001).

## 6 CONCLUSIONS

We have carried out a comprehensive survey of the compositions of hot H-rich DA white dwarfs based on archival data acquired by the *FUSE* mission. The sample includes 89 stars whose spectra are of high enough signal-to-noise to allow abundance measurements with sensible uncertainties to be made, more than a factor 3 larger than previous studies with *IUE* and *HST* spectra (Barstow et al. 2003b). The *FUSE* waveband gives access to abundances of C and Si examined by Barstow et al (2003b), but adds data for P and S, which are not available in any other waveband. As well as increasing the number of stars studied, this sample also has a more uniform coverage of the DA white dwarf temperature range above 16000K. It is also relatively unbiased in atmospheric composition as most of the targets were originally selected for ISM studies. Hence, it should be representative of the general DA white dwarf population. From this work it can be firmly established that a minimum of 23 per cent of DA stars show detectable levels of heavy elements, a fraction that increases with effective temperature.

It is interesting that, while radiative forces are expected to play a significant role in determining the element abundances in white dwarf envelopes, the abundance patterns seen in this sample bear little relation to what we would expect. First, we see stars with apparently pure H atmospheres and envelopes containing heavy elements across the whole temperature range. Second, where a particular element is detected, there is considerable scatter in the measured abundances within any particular temperature bin: generally larger than can be accounted for by the range of white dwarf gravities. Third, the radiative levitation calculations predict strong temperature dependences for the abundances of the elements studied in this paper, which are not seen. The most striking outcome of the comparison between predicted and measured abundances in Figures 6-10 is this: at temperatures where radiative levitation predicts significant elemental abundances, measured values are almost always lower, apart from the narrow temperature range where the levitation effect "switches on"; but, where no measureable abundances are expected, there are stars whose atmospheres do contain material at abundance levels similar to those seen across the temperature range. The main conclusion to be drawn from these results is that the observed abundances are dominated by the supply of material in the atmosphere. For example, when radiative levitation is "on" there is insufficient material in the atmospheric reservoir to supply the predicted equilibrium abundances. Furthermore, the presence of photospheric heavy elements when radiative levitation is "off" suggests that, for these stars at least, the envelope is being supplied from an external source.

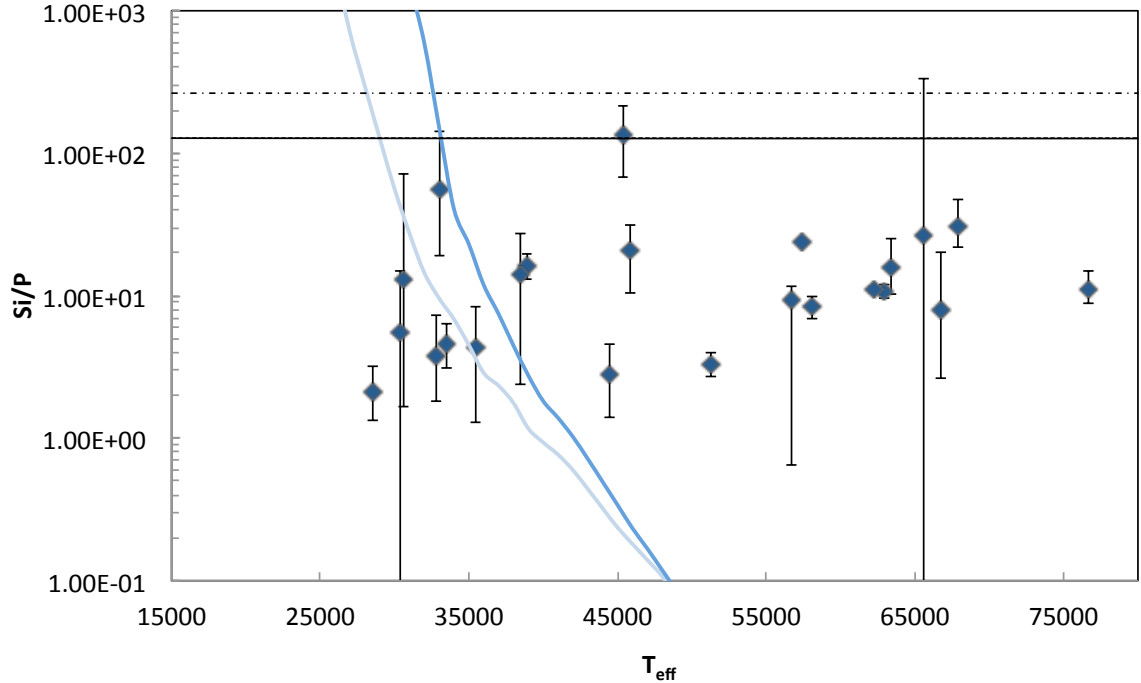

**Figure 12.** The ratio of Si:P abundance (by number) for the white dwarfs where both elements are detected, as a function of $T_{eff}$. The solid curves and horizontal lines are as in Figure 9. However, the CI chondrite ratio is almost identical to the adopted Solar abundance and is not easily visible in the plot.

   As discussed in the introduction, there is strong evidence that extra-solar planetary debris is being accreted by a substantial fraction of white dwarfs. The evidence for this arises from the presence of heavy elements in the atmospheres of these cooler objects, where radiative levitation cannot be a significant affect, and the relative abundances of material, which are similar to rocky elements of our Solar System. In our analysis of the hotter stars, carbon is similarly depleted, compared to silicon. Therefore, rocky extra-solar planetary debris is also a plausible source of the material we observe. It is well established that the residence times for accreted material are very short (~days) in the absence of radiative support. Therefore, in the cooler white dwarfs, the presence of detectable heavy elements indicates that either accretion is continuous or there must have been an accretion event shortly before the observation was made. This is consistent with the very long lifetimes predicted for the dusty disks that are believed to provide the source of material. In our sample, accretion is most likely to be from short-lived gas disks. However, as radiative levitation should operate over a significant part of the temperature range, any accreted material will remain in the atmosphere so long as the radiative support is sufficient. Therefore, the shorter lifetime of the supply of material does not present a problem to our hypothesis.

   Radiative levitation has long been accepted as the explanation of the ubiquitous heavy elements found in hot DA white dwarf atmospheres. It has also been assumed, perhaps without much careful examination, that the reservoir of material was internal, left in the envelope following evolution through the AGB. However, this model was unable to explain anomalies such as the pure H envelope of the 50000 K white dwarf HZ43 and the differences in composition found between stars with similar temperature and gravity. Based on the evidence presented here, we propose an alternative model, where the reservoir of heavy elements is external to the white dwarf, probably extra-solar planetary debris. Material is accreted into the atmosphere from this source and is retained in the envelope through radiative levitation, which then moderates the abundance as its effect declines or increases according to stellar temperature and the atomic cross section of specific ions. This can explain the diversity of compositions. White dwarfs with pure H envelopes would not have any source of debris, or at least so little that accretion is too low level to detect. The range of observed abundances would arise from differences in the mass of accreted material or frequency of accretion events or accretion rates; abundances will tend to be cumulative while radiation pressure is significant but may also decline as stars cool and the pressure decreases. Finally, since accretion can continue beyond the temperature at which radiative levitation becomes ineffective, the cooler stars can still exhibit photospheric heavy elements, if there is debris present. Recent work by Chayer (2013), examining the effect of radiative levitation and accretion in DA white dwarfs at temperatures below the range of this sample, provides some support for our hypothesis. Deal et al. (2013) have also studied the effects of accretion in cooler DA and DB white dwarfs. We would encourage such calculations to be extended to the higher temperatures covered by this paper.

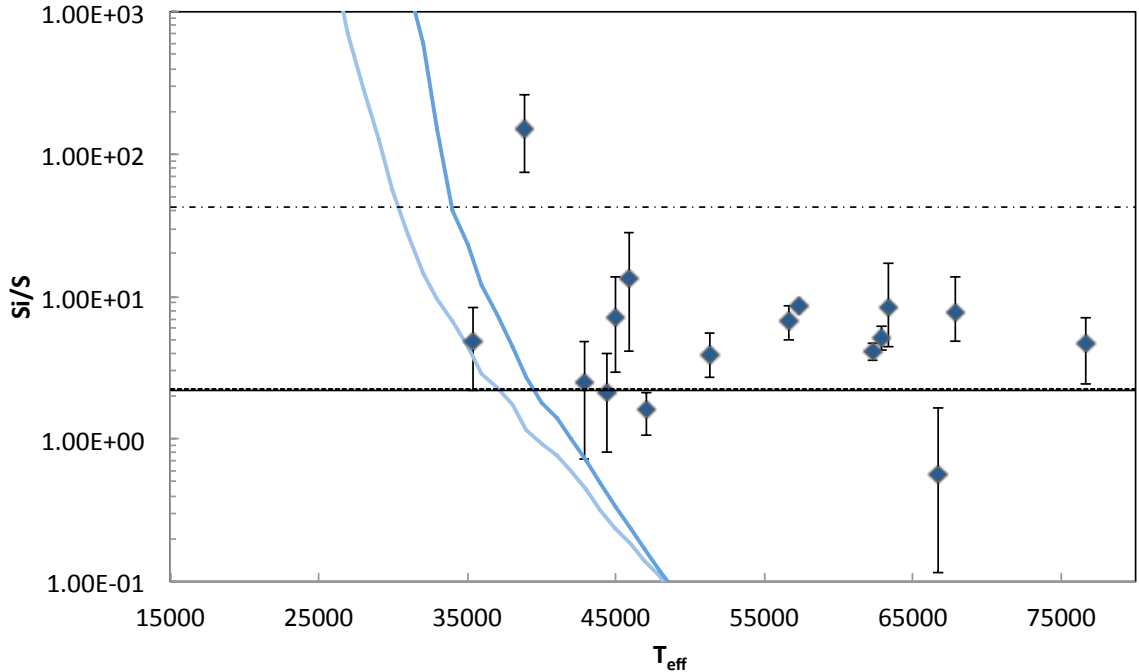

**Figure 13.** The ratio of Si:S abundance (by number) for the white dwarfs where both elements are detected, as a function of $T_{eff}$. The solid curves and horizontal lines are as in Figure 9. However, the CI chondrite ratio is almost identical to the adopted Solar abundance and is only just visible in the plot.

There are additional observational tests that could be carried out. The only strong indicator of the composition of the supposed accreted material in the *FUSE* spectra is the Si:C ratio. Extending the observations of the sample to the longer wavelengths covered by *HST* would give access to N, O, Fe and Ni features and a more detailed compositional analysis. In a significant fraction of stars the abundance measurements have quite large errors, which limits the potential for detailed object-by-object comparisons. The high sensitivity of the COS spectrograph on *HST*, can provide much better signal-to-noise data than is available in the *FUSE* archive. Finally, if accretion is the supply mechanism for the detected heavy elements, it is might not be constant in time.

**ACKNOWLEDGEMENTS**

The work reported in this paper was based on observations made with the *FUSE* observatory. Extensive use was made of the Mikulski Archive for Space Telescopes (MAST). MAB and JKB are supported by STFC, UK. SLC acknowledges support from the College of Science and Engineering at the University of Leicester. JBH and IH wish to acknowledge support provided by NASA through grant NNG056GC46G